\documentclass[twocolumn]{aastex7}
\usepackage{hyperref}
\defcitealias{akins_strong_2025}{(1)}
\defcitealias{arevalo-gonzalez_agn_2026}{(2)}

\newcommand{\jwst}{\textit{JWST}}
\newcommand{\mbh}{$M_{\rm BH}$}
\newcommand{\mstar}{$M_{*}$}
\newcommand{\mbhmstar}{$M_{\rm BH}$/$M_{*}$}
\newcommand{\beagle}{\textsc{beagle-agn}}

\newcommand{\rion}[2]{{\ensuremath{\mbox{\rm #1$\,${\sc\expandafter{\romannumeral#2\relax}}}}}}
\newcommand{\HII}{\rion{H}{2}}
\newcommand{\Ha}{\ensuremath{\mbox{\rm H$\alpha$}}}
\newcommand{\Hb}{\ensuremath{\mbox{\rm H$\beta$}}}
\newcommand{\OIII}{[\rion{O}{3}]}
\newcommand{\NII}{[\rion{N}{2}]}
\newcommand{\SII}{[\rion{S}{2}]}

\newcommand{\Msun}{${\rm M}_{\odot}$}
\newcommand{\z}{\textit{z}}

\begin{document}

\title{A Rapid Evolution in the Observed $M_{\rm BH}$/$M_{*}$ Relation at $z$ $>$ 3 Revealed via Spectro-photometric SED-Modeling}

\correspondingauthor{Ansh R. Gupta \\ Email: \href{anshrg@utexas.edu}{anshrg@utexas.edu}}
\author[0000-0003-4242-8606]{Ansh R. Gupta}
\altaffiliation{NSF Graduate Research Fellow}
\affiliation{Department of Astronomy, The University of Texas at Austin, 2515 Speedway Boulevard, Austin, TX 78712, USA}
\affiliation{Cosmic Frontier Center, The University of Texas at Austin, Austin, TX 78712, USA}
\email[]{anshrg@utexas.edu}  

\author[0000-0003-1282-7454]{Anthony Taylor}
\affiliation{Department of Astronomy, The University of Texas at Austin, 2515 Speedway Boulevard, Austin, TX 78712, USA}
\affiliation{Cosmic Frontier Center, The University of Texas at Austin, Austin, TX 78712, USA}
\email{anthony.taylor@austin.utexas.edu}

\author[0000-0002-9551-0534]{Emma Curtis-Lake}
\affiliation{Centre for Astrophysics Research, Department of Physics, Astronomy and Mathematics, University of Hertfordshire, Hatfield AL10 9AB, UK}
\email{e.curtis-lake@herts.ac.uk}

\author[0009-0002-0651-5761]{Maddie Silcock}
\affiliation{Centre for Astrophysics Research, Department of Physics, Astronomy and Mathematics, University of Hertfordshire, Hatfield AL10 9AB, UK}
\email{m.s.silcock@herts.ac.uk}

\author[0000-0003-2332-5505]{\'{O}scar A. Ch\'{a}vez Ortiz}
\altaffiliation{NASA FINESST Fellow}
\affiliation{Department of Astronomy, The University of Texas at Austin, 2515 Speedway Boulevard, Austin, TX 78712, USA}
\affiliation{Cosmic Frontier Center, The University of Texas at Austin, Austin, TX 78712, USA}
\email{chavezoscar009@utexas.edu}

\author[0000-0001-8519-1130]{Steven L. Finkelstein}
\affiliation{Department of Astronomy, The University of Texas at Austin, 2515 Speedway Boulevard, Austin, TX 78712, USA}
\affiliation{Cosmic Frontier Center, The University of Texas at Austin, Austin, TX 78712, USA}
\email{stevenf@astro.as.utexas.edu}

\author[0000-0003-3596-8794]{Hollis B. Akins}
\altaffiliation{NSF Graduate Research Fellow}
\affiliation{Department of Astronomy, The University of Texas at Austin, 2515 Speedway Boulevard, Austin, TX 78712, USA}
\affiliation{Cosmic Frontier Center, The University of Texas at Austin, Austin, TX 78712, USA}
\email{hollis.akins@utexas.edu}

\author[0000-0001-8534-7502]{Bren E. Backhaus}
\affil{Department of Physics and Astronomy, University of Kansas, Lawrence, KS 66045, USA}
\email{bren.backhaus@ku.edu}

\author[0000-0001-6813-875X]{Guillermo Barro}
\affiliation{University of the Pacific, Stockton, CA 90340 USA}
\email{gbarro@pacific.edu}

\author[0000-0003-0492-4924]{Laura Bisigello}
\affiliation{INAF-Osservatorio Astronomico di Padova, Via dell’Osservatorio 5, 35122 Padova, Italy}
\email{laura.bisigello@inaf.it}

\author[0000-0001-5384-3616]{Madisyn Brooks}
\altaffiliation{NSF Graduate Research Fellow}
\affil{Department of Physics, 196A Auditorium Road, Unit 3046, University of Connecticut, Storrs, CT 06269, USA}
\email{madisyn.brooks@uconn.edu}

\author[0000-0002-0930-6466]{Caitlin M. Casey}
\affiliation{Department of Physics, University of California, Santa Barbara, Santa Barbara, CA 93106, USA}
\affiliation{Cosmic Dawn Center (DAWN), Denmark}
\email{cmcasey@ucsb.edu}

\author[0000-0003-3458-2275]{Stephane Charlot}
\affiliation{Sorbonne Universit\'e, CNRS, UMR 7095, Institut d'Astrophysique de Paris, 98 bis bd Arago, 75014 Paris, France}
\email{charlot@iap.fr}

\author[0000-0002-7636-0534]{Jacopo Chevallard}
\affiliation{Department of Physics, University of Oxford, Denys Wilkinson Building, Keble Road, Oxford OX1 3RH, UK}
\email{jacopo.chevallard@physics.ox.ac.uk}

\author[0000-0001-6865-2871]{Anna Feltre}
\affiliation{INAF—Osservatorio Astrofisico di Arcetri, Largo E. Fermi 5, I-50125, Firenze, Italy}
\email{anna.feltre@inaf.it}

\author[0000-0003-3248-5666]{Giovanni Gandolfi}
\affiliation{INAF – Osservatorio Astronomico di Roma, via Frascati 33, 00078, Monteporzio Catone, Italy}
\email{giovanni.gandolfi@inaf.it}

\author[0000-0002-7831-8751]{Mauro Giavalisco}
\affiliation{University of Massachusetts Amherst, 710 North Pleasant Street, Amherst, MA 01003-9305, USA}
\email{mauro@umass.edu}

\author[0000-0001-9440-8872]{Norman A. Grogin}
\affiliation{Space Telescope Science Institute, 3700 San Martin Drive, Baltimore, MD 21218, USA}
\email{nagrogin@stsci.edu}

\author[0000-0002-3301-3321]{Michaela Hirschmann}
\affiliation{Institute of Physics, Laboratory for Galaxy Evolution, EPFL, Observatoire de Sauverny, Chemin Pegasi 51, CH-1290 Versoix, Switzerland}
\email{michaela.hirschmann@epfl.ch}

\author[0000-0003-4512-8705]{Tiger Yu-Yang Hsiao}
\affiliation{Department of Astronomy, The University of Texas at Austin, 2515 Speedway Boulevard, Austin, TX 78712, USA}
\affiliation{Cosmic Frontier Center, The University of Texas at Austin, Austin, TX 78712, USA}
\email{tiger.hsiao@utexas.edu}

\author[0000-0003-4242-8606]{Junehyoung Jeon}
\altaffiliation{NSF Graduate Research Fellow}
\affiliation{Department of Astronomy, The University of Texas at Austin, 2515 Speedway Boulevard, Austin, TX 78712, USA}
\affiliation{Cosmic Frontier Center, The University of Texas at Austin, Austin, TX 78712, USA}
\email{junehyoung.jeon@austin.utexas.edu}

\author[0000-0002-1590-0568]{Shardha Jogee}
\affiliation{Department of Astronomy, The University of Texas at Austin, 2515 Speedway Boulevard, Austin, TX 78712, USA}
\affiliation{Cosmic Frontier Center, The University of Texas at Austin, Austin, TX 78712, USA}
\email{sj@astro.as.utexas.edu}

\author[0000-0001-9187-3605]{Jeyhan S. Kartaltepe}
\affil{Laboratory for Multiwavelength Astrophysics, School of Physics and Astronomy, Rochester Institute of Technology, 84 Lomb Memorial Drive, Rochester, NY 14623, USA}
\email{jsksps@rit.edu}

\author[0000-0002-8360-3880]{Dale D. Kocevski}
\affiliation{Department of Physics and Astronomy, Colby College, Waterville, ME 04901, USA}
\email{dkocevsk@colby.edu}

\author[0000-0002-6610-2048]{Anton M. Koekemoer}
\affiliation{Space Telescope Science Institute, 3700 San Martin Drive,
Baltimore, MD 21218, USA}
\email{koekemoer@stsci.edu}

\author[0000-0002-5588-9156]{Vasily Kokorev}
\affiliation{Department of Astronomy, The University of Texas at Austin, 2515 Speedway Boulevard, Austin, TX 78712, USA}
\affiliation{Cosmic Frontier Center, The University of Texas at Austin, Austin, TX 78712, USA}
\email{vasily.kokorev.astro@gmail.com}

\author[0000-0002-9393-6507]{Gene C. K. Leung}
\affiliation{MIT Kavli Institute for Astrophysics and Space Research, 77 Massachusetts Ave., Cambridge, MA 02139, USA}
\email{gckleung@mit.edu}

\author[0000-0003-1581-7825]{Ray A. Lucas}
\affiliation{Space Telescope Science Institute, 3700 San Martin Drive, Baltimore, MD 21218, USA}
\email{lucas@stsci.edu}

\author[0000-0001-9879-7780]{Fabio Pacucci}
\affiliation{Center for Astrophysics $\vert$ Harvard \& Smithsonian, Cambridge, MA 02138, USA}
\affiliation{Black Hole Initiative, Harvard University, Cambridge, MA 02138, USA}
\email{fabio.pacucci@cfa.harvard.edu}

\author[0000-0003-3382-5941]{Nor Pirzkal}
\affiliation{ESA/AURA}
\email{npirzkal@stsci.edu}

\author[0000-0003-0390-0656]{Adele Plat}
\affiliation{Institute of Physics, Laboratory for Galaxy Evolution, EPFL, Observatoire de Sauverny, Chemin Pegasi 51, CH-1290 Versoix, Switzerland}
\email{adele.plat@epfl.ch}

\author[0000-0002-6748-6821]{Rachel S. Somerville}
\affiliation{Center for Computational Astrophysics, Flatiron Institute, 162 5th Avenue, New York, NY 10010, USA}
\email{rsomerville@flatironinstitute.org}

\author[0000-0002-1410-0470]{Jonathan R. Trump}
\affiliation{Department of Physics, 196A Auditorium Road, Unit 3046, University of Connecticut, Storrs, CT 06269, USA}
\email{jonathan.trump@uconn.edu}

\author[0000-0002-5268-2221]{Alba Vidal-Garc\'ia }
\affiliation{Observatorio Astronómico Nacional, C/ Alfonso XII 3, 28014 Madrid, Spain}
\email{vidal@iap.fr}

\author[0000-0002-9373-3865]{Xin Wang}
\affiliation{School of Astronomy and Space Science, University of Chinese Academy of Sciences (UCAS), Beijing 100049, China}
\affiliation{National Astronomical Observatories, Chinese Academy of Sciences, Beijing 100101, China}
\affiliation{Institute for Frontiers in Astronomy and Astrophysics, Beijing Normal University, Beijing 102206, China}
\email{xwang@ucas.ac.cn}

\author[0000-0003-3466-035X]{{L. Y. Aaron} {Yung}}
\affiliation{Space Telescope Science Institute, 3700 San Martin Drive, Baltimore, MD 21218, USA}
\email{yung@stsci.edu}

\begin{abstract}

Spectroscopic observations from \textit{JWST} have uncovered a plethora of active galactic nuclei (AGN) at $z$\,$\gtrsim$\,4 with black hole (BH) mass ($M_{\rm BH}$) to stellar mass ($M_{*}$) ratios significantly above the local relation when using standard virial mass scaling relations. However, $M_{*}$\ estimates of AGN may be inaccurate due to limitations in spectral energy distribution (SED) fitting codes, exemplified by a lack of physically-motivated AGN line emission models. Here, we fit NIRSpec/PRISM spectra of 39 galaxies at $z$\,$\sim$\,3.5--7 selected as broad-line AGN from the CEERS and RUBIES surveys. Applying kinematic decompositions from NIRSpec/G395M spectra, we fit their continuum and narrow-component line fluxes using the \textsc{beagle-agn}\ SED fitting tool. While limitations of \textsc{beagle-agn}\ make it difficult to model little red dots (LRDs), we find that $M_{*}$\ estimates of non-LRDs are, surprisingly, only modestly impacted by the inclusion or not of AGN narrow-line region (NLR) and continuum emission model components. We further find that non-LRD AGN at $z$\,$\lesssim$\,3.5 are consistent with the local $M_{\rm BH}$/$M_{*}$\ relation while those at $z$\,$\gtrsim$\,4.5 display elevated ratios. While we cannot rule out observational biases or systematic uncertainties as partial causes, this transition over just $\sim$\,500\,Myr is driven entirely by changes in $M_{*}$\ rather than an evolving $M_{\rm BH}$\ distribution. These findings are consistent with models in which rapid BH growth results in elevated $M_{\rm BH}$/$M_{*}$\ ratios at early times, with a swift late-time assembly of host galaxies returning sources to the local relation at $z$\,$<$\,4.
\end{abstract}


\section{Introduction}\label{sec:1}
 
Many high-redshift active galactic nuclei (AGN) have central black holes (BHs) with uncomfortably high masses (\mbh) from standard single epoch virial scaling relations. Growing BHs to log(\mbh/\Msun)\,$\sim$\,7--9 within a Gyr of the Big Bang presents a serious challenge for models of BH seeding and growth. This was first recognized following ground-based studies of \z\,$>$\,6 quasars \citep[e.g.][]{inayoshi_assembly_2020, yung_semianalytic_2021, pacucci_search_2022, fan_quasars_2023} and has been further exacerbated by observations from \jwst\ \citep[][]{gardner_james_2006, rigby_science_2023, gardner_james_2023}, which has uncovered an abundant population of low-to-intermediate luminosity AGN at \z\,$\gtrsim$\,4 \citep[$L_{\rm bol}$\,$\sim$\,10$^{43}$--$10^{45}$\,erg\,s$^{-1}$; e.g.][]{kocevski_hidden_2023, harikane_jwst_2023, goulding_uncover_2023, kovacs_candidate_2024, maiolino_small_2024, juodzbalis_jades_2026}. One major outstanding question is whether these early BHs have masses which are high only in absolute terms or also relative to the stellar masses (\mstar) of their host galaxies when compared to AGN in the local universe, which follow a well-characterized relation with log(\mbhmstar)\,$\sim$\,$-3$ \citep{kormendy_coevolution_2013,reines_relations_2015,greene_intermediatemass_2020}. AGN out to \z\,$\sim$\,3 appear to have values of \mbhmstar\ consistent with the local relation \citep{suh_no_2020, mountrichas_coevolution_2023, sun_no_2025, tanaka_mbh_2025}. However, observed AGN at $z \gtrsim 4$ appear to host ``overmassive'' BHs on a population level, with values of \mbhmstar\ significantly elevated ($\gtrsim$\,1\,dex) compared to the local relation \citep{pacucci_jwst_2023, jones_m_rm_2025} and extreme sources having published \mbhmstar\ ratios that approach or even exceed unity \citep[e.g.][]{kokorev_uncover_2023, juodzbalis_direct_2025, napolitano_dual_2025}.

Some theoretical models reproduce this trend by growing SMBHs from so-called ``heavy" BH seeds \citep[{log[\mbh/\Msun]}\,$\sim$\,4--5; e.g.][]{natarajan_first_2024, jeon_emerging_2025, pacucci_little_2025, bhowmick_heavy_2026} instead of from remnants of population III stars \citep[``light" seeds, {log[\mbh/\Msun]}\,$\sim$\,2--3; e.g.][]{madau_massive_2001, bromm_first_2004, bhowmick_introducing_2024, mehta_growth_2026}. Other models invoke widespread super-Eddington accretion, which would allow early SMBHs to rapidly grow to their observed masses and might explain enigmatic properties of high-redshift AGN such as X-ray weakness \citep[e.g.][]{madau_xray_2024, pacucci_mildly_2024, lambrides_case_2024, king_joining_2025}. Either factor could have caused SMBH growth to outpace host galaxy assembly, yielding elevated \mbhmstar\ ratios \citep[e.g.][]{bhowmick_growth_2024, hu_convergence_2025, jeon_physical_2025, prole_seedz_2026}. Gas inflows driven by major mergers or accretion from the IGM may have fueled rapid BH growth \citep[e.g.][]{capelo_growth_2015, pezzulli_sustainable_2017, trinca_seeking_2023, chon_rapid_2026}, with AGN feedback potentially suppressing star formation \citep[e.g.][]{weinberger_simulating_2017, kannan_increasing_2017, pacucci_redshift_2024}. Then, at late times, a slowdown in BH growth and reduction of AGN feedback would have allowed for a swift assembly of host galaxies \citep[e.g.][]{kokorev_silencing_2024, billand_investigating_2026}, returning sources to the local relation.

An alternate possibility is that the \textit{intrinsic} high-redshift \mbhmstar\ relation is consistent with the local relation, but the \textit{observed} \mbhmstar\ relation appears elevated due to selection effects. Current observations may only be sensitive to sources along the upper envelope of the intrinsic \mbhmstar\ relation \citep[][cf. \citealt{pacucci_jwst_2023}]{li_tip_2025, silverman_shellqsjwst_2025, sun_m_m_rm_2025, brooks_monsters_2026, geris_jades_2026}. As demonstrated by \citet{li_tip_2025}, it is the combined effects of uncertainties in \mbh\ and \mstar\ and observational biases towards detecting the most luminous AGN that may cause sources drawn from an intrinsically ``normal'' \mbhmstar\ distribution to appear elevated in \mbhmstar. Thus, understanding observational uncertainties in both \mbh\ and \mstar\ is vital.

Estimates of \mbh\ derived via bolometric luminosity extrapolations may be unreliable due to the uncertain shapes of the spectral energy distributions (SEDs) of high-redshift AGN and their wide range of Eddington ratios \citep[e.g.][]{lambrides_case_2024, leung_exploring_2025, greene_what_2026}. On the other hand, \mbh\ estimates from single-epoch (SE) virial scaling relations applied to measurements of broad Balmer emission lines \citep{greene_estimating_2005, reines_dwarf_2013} require that the locally-established relationship between the size of the broad line region (BLR) of an AGN and its luminosity \citep[the $R$--$L$ relation; e.g.][]{kaspi_relationship_2005, bentz_lowluminosity_2013} holds and that broad line widths represent motion in virial orbits. These assumptions may not be well-motivated for high-redshift AGN due to deviations from the $R$--$L$ relation under super-Eddington accretion conditions \citep[e.g.][]{du_supermassive_2018, li_reverberation_2021, abuter_dynamical_2024} and potential non-virial line broadening \citep[e.g.][]{chang_impact_2026, kokorev_deepest_2025, rusakov_jwsts_2026, brazzini_little_2026}.

While uncertainties in \mbh\ are well known, it is less clear how well \mstar\ has been measured for high-redshift AGN. \mstar\ is most commonly estimated using SED fitting codes, which are widely used to determine the basic physical properties of galaxies \citep[e.g.][]{conroy_modeling_2013, pacifici_art_2023, iyer_spectral_2025}. However, applying these codes to AGN has been challenging for multiple reasons. First, AGN can produce strong continuum emission which can equal or strongly outshine stellar light from the host galaxy. Modeling this continuum is highly non-trivial, though some SED fitting codes are beginning to implement such models \citep[e.g.][]{boquien_cigale_2019, martinez-ramirez_agnfitterrx_2024, buchner_genuine_2024, lovell_synthesizer_2025} and attempts have been made to apply these to high-redshift AGN \citep[e.g.][]{florez_exploring_2020, maiolino_jades_2024, leung_exploring_2025, ronayne_mega_2025, juodzbalis_jades_2026}. It has been particularly challenging to use SED fitting codes to model little red dots \citep[LRDs; e.g.][]{matthee_little_2024, greene_uncover_2024, kokorev_census_2024, kocevski_rise_2025, akins_cosmosweb_2025} due to the uncertain nature of their continuum emission. Explanations for their rest-optical continuum range from central engines with light reprocessed through extremely dense gas \citep[e.g.][]{inayoshi_extremely_2025, ji_blackthunder_2025, naidu_black_2025, degraaff_remarkable_2025} to supermassive stars \citep[e.g.][]{zwick_little_2025, nandal_supermassive_2026, chisholm_little_2026}. Their rest-UV emission may arise from stellar emission \citep[e.g.][]{chen_host_2025, chen_zsimeq04_2026, zhuang_nexus_2026} or an accreting BH \citep[e.g.][]{greene_uncover_2024, li_little_2025, pacucci_little_2026}. 

While these results of modeling AGN continuum emission are promising, interpreting the emission lines of AGN has remained a significant challenge. Strong emission lines may contain light from the host galaxy, the BLR, and the narrow line region \citep[NLR; e.g.][]{scholtz_jades_2025, mazzolari_narrow_2025, davis_extreme_2026}. Since most SED fitting codes do not model the BLR or NLR, previous works mask emission lines when performing SED fitting of AGN \citep[or marginalize over them using analytic profiles, e.g.][]{wang_rubies_2024}. Works analyzing low-\z\ quasars (\z\,$\lesssim$\,1) achieve high-quality AGN and host galaxy decomposition by fitting spectra (including emission lines) using linear combinations of quasar and galaxy templates \citep[e.g.][]{vandenberk_spectral_2006, guo_pyqsofit_2018, ilic_fantastic_2023, ren_priorinformed_2024}. However, this technique has generally not been fruitful for modeling high-redshift AGN (especially LRDs), whose emission line and host galaxy properties appear quite different from their low-\z\ counterparts \citep[e.g.][]{ma_uncover_2025, maiolino_jwst_2025, taylor_caperslrdz9_2025}.

Here, we utilize a version of the BayEsian Analysis of GaLaxy sEds \citep[\textsc{beagle};][]{chevallard_modelling_2016} SED fitting code which includes NLR and AGN continuum emission models \citep[\beagle;][]{vidal-garcia_beagleagn_2024, silcock_characterising_2025, chavezortiz_significant_2025}. We use this code to fit continuum emission and narrow-component emission line fluxes of 39 galaxies (including 16 LRDs) at \z\,$\sim$\,3.5--7 which have NIRSpec/PRISM+G395M spectra and are selected as broad-line (BL) AGN, with an aim to estimate more reliable \mstar\ values. We test the effectiveness of these new AGN emission models in reproducing the observed SEDs of our sample and examine the effects of their inclusion or not on \mstar\ estimates. Finally, we use these new \mstar\ values to examine the evolution of the observed \mbhmstar\ relation over cosmic time.

This paper is organized as follows. We introduce our data sources and parent sample in Section \ref{sec:2}. In Section \ref{sec:3}, we describe our sample selection and fitting methodology. Section \ref{sec:4} presents our main results, and we interpret these and discuss relevant limitations in Section \ref{sec:5}. Lastly, we summarize our work in Section \ref{sec:6}. Throughout this work, we use a $\Lambda$CDM cosmology with H$_0$\,=\,70\,km\,s$^{-1}$\,Mpc$^{-1}$, $\Omega_m$\,=\,0.3, and $\Omega_\Lambda$\,=\,0.7. All magnitudes are reported in the AB system \citep{oke_secondary_1983}.

\section{Data}\label{sec:2}
In this work, we analyze 39 of the 62 BL AGN selected by \citet[][see Section \ref{sec:3.1} for details on sample selection]{taylor_broadline_2025} via the detection of broad \Ha\ emission in \jwst/NIRSpec G395M spectroscopy from the Cosmic Evolution Early Release Science (CEERS) survey \citep[ERS\#1345, PI: Finkelstein;][]{finkelstein_cosmic_2025} and the Red Unknowns: Bright Infrared Extragalactic Survey \citep[RUBIES, Cycle 2 GO\#4233, PIs: de Graaf and Brammer;][]{degraaff_rubies_2025}. We utilize this sample as it spans a wide range in \mbh\ and redshift (Figure \ref{fig:mbh-z}) and represents the largest published sample of \jwst-detected BL AGN in this redshift range which currently do not have published spectroscopic \mstar\ measurements.

\subsection{Photometry}
To account for slit losses and obtain accurate \mstar\ estimates from fitting, our spectra must be rescaled to match photometric measurements. Thus, accurate photometry is essential for our science goals.

We use \jwst/NIRCam imaging and measured photometry from CEERS in the Extended Groth Strip \citep[EGS;][]{davis_allwavelength_2007} field (hereafter the CEERS field) and from the Public Release IMaging for Extragalactic Research (PRIMER) survey (Cycle 1 GO\#1837, PI: Dunlop; Dunlop et al.\ in prep.) in the Ultra-deep Survey (UDS) field.

In the CEERS field, we use reduced v1.0 imaging from the CEERS team \citep{bagley_ceers_2023, finkelstein_cosmic_2025} and F090W imaging over a similar footprint from an additional \jwst\ program (Cycle 1 GO\#2234, PI: Ba\~nados). 
We reduce the PRIMER imaging in the UDS field following a similar procedure as outlined in \citet{bagley_ceers_2023}. 

Both fields have photometric coverage in the NIRCam F090W, F115W, F150W, F200W, F277W, F356W, F410M, and F444W filters. We additionally use \textit{HST}/ACS F606W and F814W coverage over both fields from the Cosmic Assembly Near-infrared Deep Extragalactic Legacy Survey \citep[CANDELS;][]{koekemoer_candels_2011,grogin_candels_2011}. In both fields, we leverage the UNICORN catalog (Finkelstein et al.\ in prep.), which contains uniformly processed NIRCam and HST photometric measurements over major \jwst\ legacy fields. The methodology used to produce the UNICORN catalog is essentially the same as the procedure described in \citet{finkelstein_complete_2024}, including additional updates as described in \citet{taylor_broadline_2025}, with a focus on accurate colors, total flux estimates, and flux uncertainties.

\subsection{Spectroscopy}
We are interested in performing spectro-photometric modeling to obtain \mstar\ estimates, so we focus on the existing PRISM spectroscopy covering these sources. For detailed descriptions of the G395M data reduction procedures and initial sample selection, we refer the reader to \citet{taylor_broadline_2025}.

We use PRISM spectroscopy from CEERS \citep{arrabalharo_spectroscopic_2023} and RUBIES \citep{degraaff_rubies_2025}. The CEERS program obtained spectra using the \jwst/NIRSpec PRISM/CLEAR disperser/filter combination over six pointings. The RUBIES program obtained PRISM/CLEAR spectra across six pointings in the CEERS field and 12 pointings in the UDS field.

Our data reduction procedure largely follows default steps from the JWST Data Calibration Pipeline, with some modifications to the artifact-rejection procedure. 1D spectra are optimally extracted from 2D spectra following the methodology of \citet{horne_optimal_1986}. The pipeline flux error spectra are scaled up by a conservative factor of 1.7$\times$ in order to match empirical error arrays measured from the standard deviation of the science flux spectrum. Finally, flux spectra and error arrays are rescaled to match the photometry. In this rescaling, we simulate photometric measurements for each source using JWST filter transmission curves and the observed PRISM spectrum. We then fit a third-order Chebyshev polynomial to the ratios of the measured NIRCam photometry to the PRISM-derived photometry. The PRISM spectrum is finally multiplied by the resulting fitted curve to produce a photometrically calibrated spectrum. We repeat this process for each spectrum individually and use the resulting calibrated spectra for the remainder of this work. In principle, a photometric aperture and a spectroscopic slit aperture may sample somewhat different mixes of stars and NLR emission, which could cause emission lines to be over- or under-weighted in the final spectrum. We discuss the implications of this in Section \ref{sec:5.2}.

\section{Methods}\label{sec:3}
\subsection{Sample Selection}\label{sec:3.1}
In this work, we analyze 39 out of a total of 62 BL AGN from the CEERS and RUBIES surveys selected using G395M spectroscopy in \citet{taylor_broadline_2025}, with sample selection described as follows. We first adopt their entire sample as our parent sample. Since we perform SED fitting on the PRISM spectra of these sources, we start by limiting our sample to the subset of their published sources which have PRISM spectra without significant contamination or data quality issues. We also discard any sources without NIRCam coverage, as our spectral flux calibration relies on photometric measurements. Next, we require the PRISM spectrum of each source to cover both the \Ha\ and \Hb\ emission lines and the rest-frame UV continuum down to at least 1500\,\AA, as these are the minimum set of features which allow us to probe the relative contributions of stellar and AGN light in the spectrum (see Section \ref{sec:5.2}). We exclude two remaining sources, RUBIES-UDS-154183 and RUBIES-UDS-155916, from our analysis for the following reasons. RUBIES-UDS-154183 (also referred to as ``The Cliff'') is an LRD with an extremely strong Balmer break that is well-fitted only by models \citep[e.g.][]{kido_black_2025} which invoke the effects of dense gas around a central engine \citep{degraaff_remarkable_2025}. \beagle\ does not currently include such models, so we are unable to reproduce the rest-optical emission of this source and we thus exclude it from our final sample. RUBIES-UDS-155916 is a quiescent galaxy \citep{zhang_rubies_2026} with strong absorption in Balmer series lines (as seen in its PRISM spectrum); due to this, its measured narrow \Ha\ flux is completely degenerate with a fitted absorption component  (even in the G395M spectrum). Since our SED fitting relies on accurate narrow-line fluxes (as we describe in Section \ref{sec:3.3}), we also exclude this source from our final analysis. After these selections, we are left with a sample of 39 BL AGN which are analyzed in the remainder of this work.

\begin{figure}[t!]
    \centering
    \includegraphics[width=\linewidth]{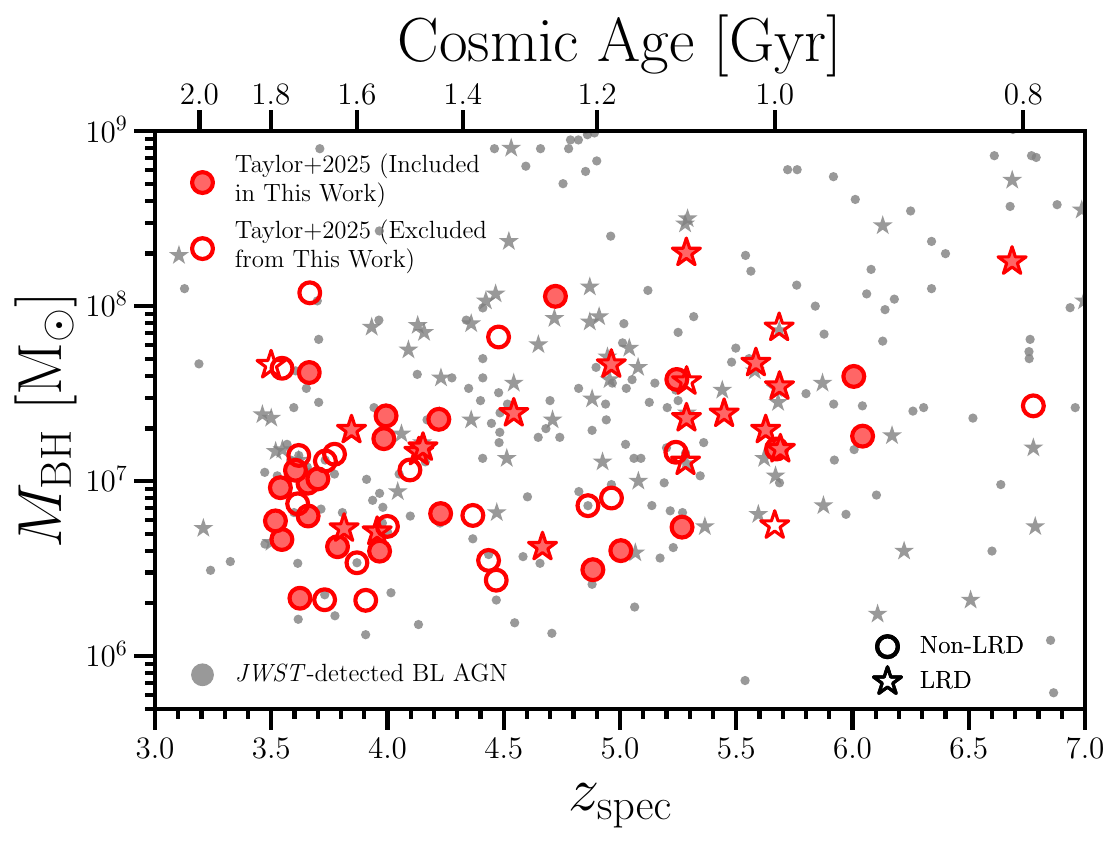}
    \caption{Black hole mass vs.\ spectroscopic redshift for \jwst-detected BL AGN. Sources published in \citet{taylor_broadline_2025} are shown as red symbols, with filled markers indicating the subset analyzed in this work and empty markers representing those excluded from our analysis. Other \jwst-detected BL AGN published in previous works (listed in Appendix \ref{apdx:c}) are shown in gray. LRDs are plotted as stars, while non-LRD BL AGN are indicated with circles. Our sample constitutes a relatively representative subset of the overall high-redshift AGN sample.}
    \label{fig:mbh-z}
\end{figure}

\citet{taylor_broadline_2025} define LRDs as sources which are compact in NIRCam imaging and satisfy the continuum slope cuts of \citet{kocevski_rise_2025}. Here, we directly adopt their LRD classifications, finding that 17 of our sources are LRDs and 23 are non-LRD AGN (though our final results do not significantly change if we instead use the looser criteria of \citealt{barro_extremely_2024}). We also utilize \mbh\ estimates from \citet{taylor_broadline_2025}, which are derived from the full width at half maximum (FWHM) and line luminosity of the broad component of \Ha\ using the SE virial mass recipe of \citet{reines_dwarf_2013}. As shown in Figure \ref{fig:mbh-z}, our sample spans a wide range in both \mbh\ and redshift and comprises a representative selection of BL AGN compared to those published in previous works.

\subsection{Broad Line Decomposition}\label{sec:3.2}
\beagle\ leverages emission line fluxes from various atomic species to constrain the relative importance of stellar-driven and NLR emission (e.g., via diagnostic line ratios). Models for stellar, NLR, and AGN accretion disk continuum emission can be included, but \beagle\ does not currently support BLR emission modeling. As such, the broad components of Balmer emission lines in these spectra must be excluded. Thus, before the SED fitting, we decompose Balmer emission lines into broad and narrow components and include only the latter in the fitting. The majority of the emission lines in our sample are at best marginally resolved at the PRISM resolution. Thus, we leverage the results of \citet{taylor_broadline_2025}, who performed a decomposition of the \Ha+\NII\ line complex using G395M spectra. We adopt their ratios of broad \Ha, narrow \Ha\, and \NII\,$\lambda\lambda6550,6585$ fluxes in each source. In order to mitigate any differences in spectral flux calibration between the PRISM and G395M spectra, we apply these ratios to the total integrated flux of the \Ha+\NII\ profile from the PRISM rather than using the fluxes as directly measured in the G395M spectra.

To determine the total \Ha+\NII\ flux in each PRISM spectrum, we utilize the \textsc{emcee} Python package \citep{foreman-mackey_emcee_2019}. We model \Ha\ as a sum of two Gaussian profiles, with the intrinsic FWHM (corrected for instrumental broadening) of the narrow line restricted to $<$700\,km\,s$^{-1}$ and that of the broad line to $>$700\,km\,s$^{-1}$ (though the exact velocity cutoff is found to have minimal impact on the total \Ha\ flux). We also include single Gaussian profiles to fit each narrow emission line of the \NII\,$\lambda\lambda6550,6585$ doublet \citep[with an internal flux ratio fixed to the theoretical value of 1:2.94;][]{storey_theoretical_2000} and each component of the \SII\,$\lambda\lambda6718,6733$ doublet, with the FWHMs tied to the narrow \Ha\ line. A linear function is used to model the continuum. We fix the wavelength of each line but allow the redshift of the total profile to vary. Finally, we apply line flux ratios from the G395M spectrum to the total PRISM \Ha\ profile flux to determine the flux in each component, keeping only the narrow-component fluxes for the \beagle\ fitting (Section \ref{sec:3.3}).

\begin{figure}[t!]
    \centering
    \includegraphics[width=\linewidth]{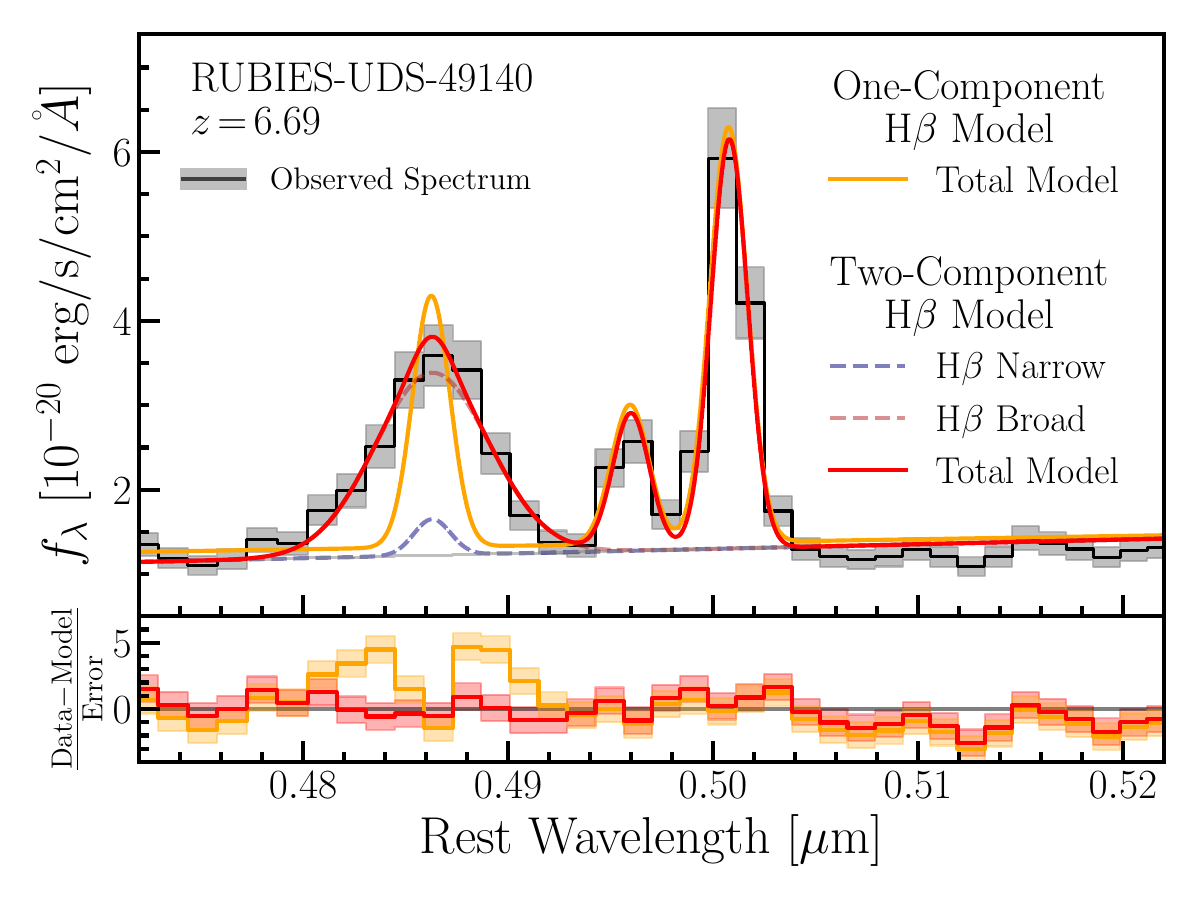}
    \caption{Best-fit one and two-component models of the \Hb+\OIII\ region of a source with significant broad \Hb. The top panel shows the total single-component fit in orange and the total two-component fit in red, with corresponding residuals in the bottom panel. Only two sources in our sample have significant broad \Hb, in qualitative agreement with the findings of \citet{brooks_here_2025}.
    }
    \label{fig:halpha-prism-fit}
\end{figure}

We next explore whether any source has significant broad \Hb\ emission that must be subtracted. 
We perform two fits to the \Hb+\OIII\ complex in each PRISM spectrum: one with only a single Gaussian component to model \Hb, and a second with both a broad and a narrow component. In both fits, we also include a Gaussian component for each line of the \OIII\,$\lambda\lambda4960,5008$ doublet \citep[with their flux ratio set to the theoretical value of 1:2.98;][]{storey_theoretical_2000}. As before, we fix the wavelength of all lines but leave the redshift of the total profile as a free parameter. We do not include a broad component when fitting \OIII\ (which can result from outflows or gas kinematics), as \citet{taylor_broadline_2025} only find evidence for broad \OIII\ flux in the G395M spectrum of a single source: RUBIES-EGS-50052. All parameters are measured completely independently from the \Ha\ emission. We compute the Bayesian information criterion \citep[BIC;][]{schwarz_estimating_1978} for each fit. To select sources with significant broad \Hb, we enforce the commonly-used criterion of BIC$_{\rm 1\,component}$\,--\,BIC$_{\rm 2\,component}$\,$>$\,$6$ \citep{kass_bayes_1995} and further require broad component S/N\,$>$\,5. Only two sources, RUBIES-EGS-49140 and RUBIES-EGS-42046, satisfy these criteria. The \Hb\ decomposition for the first of these is shown in Figure \ref{fig:halpha-prism-fit}. For these two sources, we use our PRISM spectra decomposition to measure the narrow-component \Hb\ flux. For all remaining sources (which are not found to have significant broad \Hb), we use the total \Hb\ flux from the single-Gaussian fit as the narrow-component \Hb\ flux. In principle, broad \Hb\ emission in these sources may be indistinguishable from the narrow component at the low PRISM resolution, biasing our inferred narrow \Hb\ fluxes high. However, previous results have shown that broad \Hb\ is extremely weak in high--\z\ \jwst-detected AGN, aside from a small handful of sources in which is is extremely bright and broad (likely resolvable in our PRISM spectra); excluding these few detections, broad \Hb\ remains undetected in stacks of tens of G395M spectra \citep{brooks_here_2025}.

\subsection{Fitting with BEAGLE-AGN}\label{sec:3.3}
During each step of a \beagle\ fit, a high-resolution model spectrum is convolved with the PRISM line spread function, downsampled to the wavelength grid of the observed spectrum, and compared with the observed data. To ensure we do not include broad Balmer line emission in the fitting, we mask the \Ha+\NII\ complex and \Hb+\OIII\ region in each source, choosing a uniform window which excludes these lines for all sources. We then use the ``spectral indices'' feature of \beagle, which allows us to simultaneously fit the continuum emission from the PRISM spectrum and the narrow-component emission line fluxes. In addition to the line fluxes from the Ha+\NII\ complex and \Hb+\OIII\ region, we also fit each pixel in the PRISM spectrum outside of the masked regions.

We leverage the ability of \beagle\ to simultaneously fit observed SEDs with several independent model components. For a detailed description of the most up-to-date modeling procedure in \beagle, we refer the reader to \citet{vidal-garcia_beagleagn_2024}. Here, we briefly describe each model component, their relevant fitted parameters, and our choice of priors for all fitted quantities (listed in Table \ref{tab:beagleparams}).

We begin by describing the model for stellar and associated nebular emission from \HII\ regions. Simple stellar population grids are generated using the \citet{bruzual_stellar_2003} code, specifically the updated version described in \citet{vidal-garcia_modelling_2017}. We assume a fixed \citet{chabrier_galactic_2003} initial mass function (IMF). Nebular emission is incorporated using the photoionization models of \citet{gutkin_modelling_2016}. To model dust attenuation, we apply the two-component model of \citet{charlot_simple_2000}, which includes dust in the birth clouds of newly formed stars and throughout the diffuse interstellar medium (ISM). In our models, we select a delayed-exponential star formation history (SFH), which offers both reasonable flexibility and model complexity, in addition to an independently-fitted current burst parameterized by a star formation rate (SFR) averaged over the past 10\,Myr. We select well-tested priors on SFH and other stellar parameters (including metallicity, ionization parameter, dust attenuation, etc., described in Table \ref{tab:beagleparams}) which have been successfully demonstrated in previously published works, including those studying AGN \citep[e.g.][]{sandles_bayesian_2022, curtis-lake_spectroscopic_2023, silcock_characterising_2025, chavezortiz_significant_2025}.

Importantly, in addition to modeling nebular emission from \HII\ regions, \beagle\ includes the ability to model nebular emission from the AGN NLR. Models of the NLR are constructed using the grids of \citet{feltre_nuclear_2016} with updates from \citet{mignoli_obscured_2019}. These are produced by applying the \textsc{CLOUDY} code \citep{ferland_2017_2017} to model the transfer of an AGN accretion disk spectrum (constructed as a broken power-law distinguishing the UV-optical and ionizing continua) through NLR gas. In our models, the NLR nebular emission is attenuated by the same dust component in the diffuse ISM as is applied to stellar emission (but the dust present in stellar birth clouds is not applied). The NLR grids are parameterized in terms of metallicity, ionization parameter, dust-to-metal mass ratio, gas covering fraction, and accretion disk luminosity (as a proxy for normalization). We list our choices for the prior conditions on each of these parameters in Table \ref{tab:beagleparams}. Importantly, \citet{vidal-garcia_beagleagn_2024} find that leaving all of these as free parameters introduces significant degeneracies and results in biased parameter retrieval. Following their guidance, we fix several of these quantities to the recommended values to minimize bias in inference results (see Section \ref{sec:5.2} for discussion of model assumptions). For instance, a blanket covering fraction of 10\% is assumed, and the slope of the ionizing continuum from the AGN accretion disk incident on the NLR is set to ${{\alpha}_{\rm NLR}}$\,=\,$-1.7$. This ionizing continuum slope is not the same as the UV-optical continuum slope directly observed in sources presenting an unobscured view of the accretion disk (Type I AGN). Instead, it describes the extreme UV and X-ray emission of the AGN, which is typically steeper (softer) than would be inferred from a simple extrapolation of the UV-optical continuum \citep{gierlinski_soft_2004}. 

\begin{deluxetable*}{llll}\label{tab:beagleparams}
\tablecaption{A table of allowed components in a \beagle\ fit, fitted or fixed parameters used to define each component, descriptions of each parameter, and our selected priors.}
\tablewidth{0pt}
\tablehead{
    \colhead{Component} & \colhead{Parameter} & \colhead{Description} & \colhead{Prior}
}
\startdata
Stars & $\mathrm{log}($$M_{\rm*}/\mathrm{M_{\odot}}$) & Total mass in stars & Uniform $\in [5,12]$ \\
      & $\mathrm{log}($$Z_*/\mathrm{Z}_{\odot})$ & Stellar metallicity & Uniform $\in [-2.2, 0.24]$ \\
      & $\mathrm{log}(\tau_{\rm SFR}/\mathrm{yr})$ & e-folding timescale of delayed-exponential SFH & Uniform $\in [6,12]$ \\
      & $\mathrm{log}(t_{\rm max}/\mathrm{yr})$ & Time since onset of star formation (max age of stars) & Uniform $\in [7,10.8]$ \\
      & $\mathrm{log}({\rm SFR}/\mathrm{M}_{\odot}\,\mathrm{yr}^{-1})$ & SFR over 10 Myr independent burst  & Uniform $\in [-4,4]$ \\
      & $m_{\textrm{up}}$/$\mathrm{M_{\odot}}$ & Upper mass cutoff of the \citet{chabrier_galactic_2003} IMF & Fixed to 100 \\
\hline
\HII\ regions & $\mathrm{log}($$Z_{\rm gas}^{\rm HII}/\mathrm{Z}_{\odot})$ & \HII\ region gas metallicity & Fixed to $Z_*$ \\
              & log\,$U^{\rm HII}$ & \HII\ region ionization parameter & Uniform $\in [-4,-1]$ \\
              & $\xi_d$ & \HII\ region dust-to-metal mass ratio & Fixed to 0.3 \\
              & $\hat{\tau}_{\rm V}$ & Effective V-band optical depth to stars & Exponential $\in [0,6]$ \\
              & $\mu$ & Fraction of $\hat{\tau}_V$ from dust in the diffuse ISM & Fixed to 0.4 \\
              & $n_{\rm H}^{\rm HII}$/cm$^{-3}$ & Hydrogen density in \HII\ regions & Fixed to 100 \\
\hline
AGN NLR & $\mathrm{log}($$Z_{\rm gas}^{\rm NLR}/\mathrm{Z}_{\odot})$ & NLR gas metallicity & Uniform $\in [-2.2, 0.24]$ \\
        & log\,$U^{\rm NLR}$ & NLR ionization parameter & Uniform $\in [-4,-1]$ \\
        & $\xi_d^{\rm NLR}$ & NLR dust-to-metal mass ratio & Fixed to 0.3 \\
        & $\hat{\tau}_{\rm V\!,\,NLR}$ & Effective V-band optical depth to the NLR & Fixed to $\hat{\tau}_{\rm V}$ \\
        & $\mathrm{log}$($L_{\rm acc}$) & AGN accretion disk luminosity & Uniform $\in [43,48]$ \\
        & ${{C}_{f}}$ & Covering fraction of AGN by NLR gas & Fixed to $0.1$ \\
        & ${{\alpha}_{\rm NLR}}$ & Power-law slope of ionizing radiation incident on NLR & Fixed to $-1.7$ \\
        & $n_{\rm H}^{\rm NLR}$/cm$^{-3}$ & Hydrogen density in the NLR & Fixed to 1000 \\
\hline
AGN continuum & $\mathrm{log}(f_{\rm AGN})$ & Fraction of 1500\AA\ continuum from AGN emission & Uniform $\in [-3,4]$ \\
              & ${{\alpha}_{\rm cont}}$ & Power-law slope of AGN UV-optical continuum & Fixed to $-1.54$ \\
              & $\hat{\tau}_{\rm V\!,\,cont}$ & Effective V-band optical depth to the AGN continuum & Exponential $\in [0,6]$ \\
\enddata
\end{deluxetable*}

As such, \beagle\ allows for an additional independent component to be included which models UV-optical continuum emission from the AGN accretion disk \citep[as done in e.g.][]{maiolino_jades_2024, juodzbalis_jades_2026}. The AGN continuum model is described by an analytic power-law $f_{\lambda}$\,=\,$f_{1500}(\lambda/1500$\AA)$^{{\alpha}_{\rm cont}}$, where $f_{1500}$ is the flux at 1500\AA\ and ${\alpha}_{\rm cont}$ is the UV-optical power-law continuum slope, with freely varying attenuation (independent of that of the host galaxy) modeled by an SMC dust law \citep{pei_interstellar_1992}. To determine the effects of varying ${\alpha}_{\rm cont}$, we test two distinct canonical AGN power-law continuum models which have differing values of the UV-optical slope. In one set of models, we use $\alpha_{\rm cont}$\,=\,$-1.54$, observationally derived from a large sample of SDSS quasars \citep{vandenberk_composite_2001}. In a second set of models, we set $\alpha_{\rm cont}$\,=\,$-2.33$, a theoretically derived value of the UV-optical slope of an AGN accretion disk \citep{shakura_black_1973} which we select to span a large range in $\alpha_{\rm cont}$ between the two values \citep[following][]{maiolino_jades_2024, juodzbalis_jades_2026}. In Section \ref{sec:5.2}, we explore impacts of AGN continuum modeling choices by exploring differences in fitting results between these two models and find that fits are minimally impacted by the choice of $\alpha_{\rm cont}$. Thus, we report results only for the more empirically-motivated value of $\alpha_{\rm cont}$\,=\,$-1.54$. 

In total, we perform three \beagle\ fits to each source, varying which model components are enabled to test their significance. We define these fits as follows:

\noindent \textit{i) Stellar-only} -- The model includes only stellar emission and nebular emission from \HII\ regions.

\noindent \textit{ii) Stellar+NLR} -- The model includes stellar emission, nebular emission from \HII\ regions, and nebular emission from the AGN NLR.

\noindent \textit{iii) Stellar+NLR+AGN-continuum} -- The model includes stellar emission, nebular emission from both \HII\ regions and the AGN NLR, and an AGN continuum model with an intrinsic power-law slope of $-1.54$.

For each source, we determine which of these three models best describes the observed SED using a likelihood comparison. BEAGLE performs sampling using \textsc{multinest} \citep{feroz_multinest_2009}, which computes the marginal likelihood (also referred to as global evidence, $Z$) of each model evaluation. We define the fiducial model for each source as the one which has the highest value of $Z$. We accordingly compute the Bayes factor (BF$_{12}$\,=\,$Z_1/Z_2$) between each of the three AGN models and the \textit{stellar-only} model. Using the BF offers several advantages over reduced metrics such as the BIC. On a fundamental level, comparing two models using the BF leverages the full posterior distributions, while a simple BIC test on the highest-likelihood sample from each discards the information gained from the Bayesian sampling approach. Moreover, the BF can not only produce evidence \textit{against} a given model, but it can yield evidence \textit{in support} of the null hypothesis. In our case, this key quality allows us to determine in a strict statistical sense whether any of the fitted models are strongly preferred or whether no significant difference is measured between them. Like the BIC, the BF penalizes overfitting by normalizing relative to the searched prior volume.

\subsection{Control Sample for Comparison}
Adding additional components to our models (the NLR emission and AGN continuum) might improve fits for two qualitatively different reasons. One reason is that the spectra of BL AGN differ from inactive galaxies in physically driven ways. If the \textit{stellar-only} model is unable to provide an acceptable fit to an AGN spectrum, then adding these components (which model the true physical origin of the emission) should allow our model to better reproduce the data. On the other hand, adding free parameters to a model might allow it to better match the data in an unphysical way \citep[i.e., overfitting;][]{magrisc._recovery_2015, woo_stellar_2024, jespersen_optical_2026}. While many of the components fitted by \beagle\ are modeled in a physically self-consistent way, capturing the entire complexity of galaxy physics is impossible. Our BF comparison naturally accounts for overfitting via prior-volume normalization. However, we decided to further investigate the effects of overfitting by repeating our procedure on a sample of ``control'' galaxies (defined as sources with no detected broad emission lines). If the \textit{stellar-only} models of \beagle\ are adequate to describe the spectra of control galaxies but are unable to model BL AGN spectra acceptably, then we have good evidence that a preference for AGN models among our BL AGN sample is physically meaningful and not a result of overfitting.

To select a control sample, we begin with all sources which have CEERS PRISM spectroscopy and remove known BL AGN. We bin these in observed F277W magnitude and redshift, with $\Delta m_{\rm F277W}$\,=\,$1$ and $\Delta z$\,=\,$0.5$. We randomly sample one source from each populated bin, visually inspecting the resulting spectra and removing sources with contamination or other data quality issues. To match our BL AGN sample, we also remove sources whose PRISM spectra do not cover \Ha, \Hb, and the rest-UV continuum down to 1500\,\AA. We continue resampling until one object is selected from each bin. This procedure yields a set of 23 control galaxies (with no detected broad lines) that span a similar redshift and magnitude range as our BL AGN sample. We measure the line fluxes of these sources in an equivalent manner as for our BL AGN (Section \ref{sec:3.2}). For sources which lack G395M spectral coverage, we are unable to disentangle contributions from each emission line in the \Ha+\NII\ line complex. In these cases, we assume line ratios equal to the median for sources with G395M spectroscopy. The results obtained from this comparison sample are presented in Section \ref{sec:4.1}. In principle, our selection criteria could allow some NL AGN to be included in the control sample; we discuss this and related implications in Section \ref{sec:5.3}.

\section{Results}\label{sec:4}
In this section, we present the results of \beagle\ fits to our sample. We begin by examining the results from each of the three model types, describing which are found to provide the best fits, and comparing the statistical preference for AGN models in the BL AGN sample to that of the control sample. We next examine \mstar\ estimates among non-LRDs and for LRDs, and we finally use these results to investigate the redshift-dependence of the observed \mbhmstar\ relation. Fitted parameter values and uncertainties from the best-fit model for each source are tabulated in Appendix \ref{apdx:b}.

\begin{figure*}[t!]
    \centering
    \includegraphics[width=\linewidth]{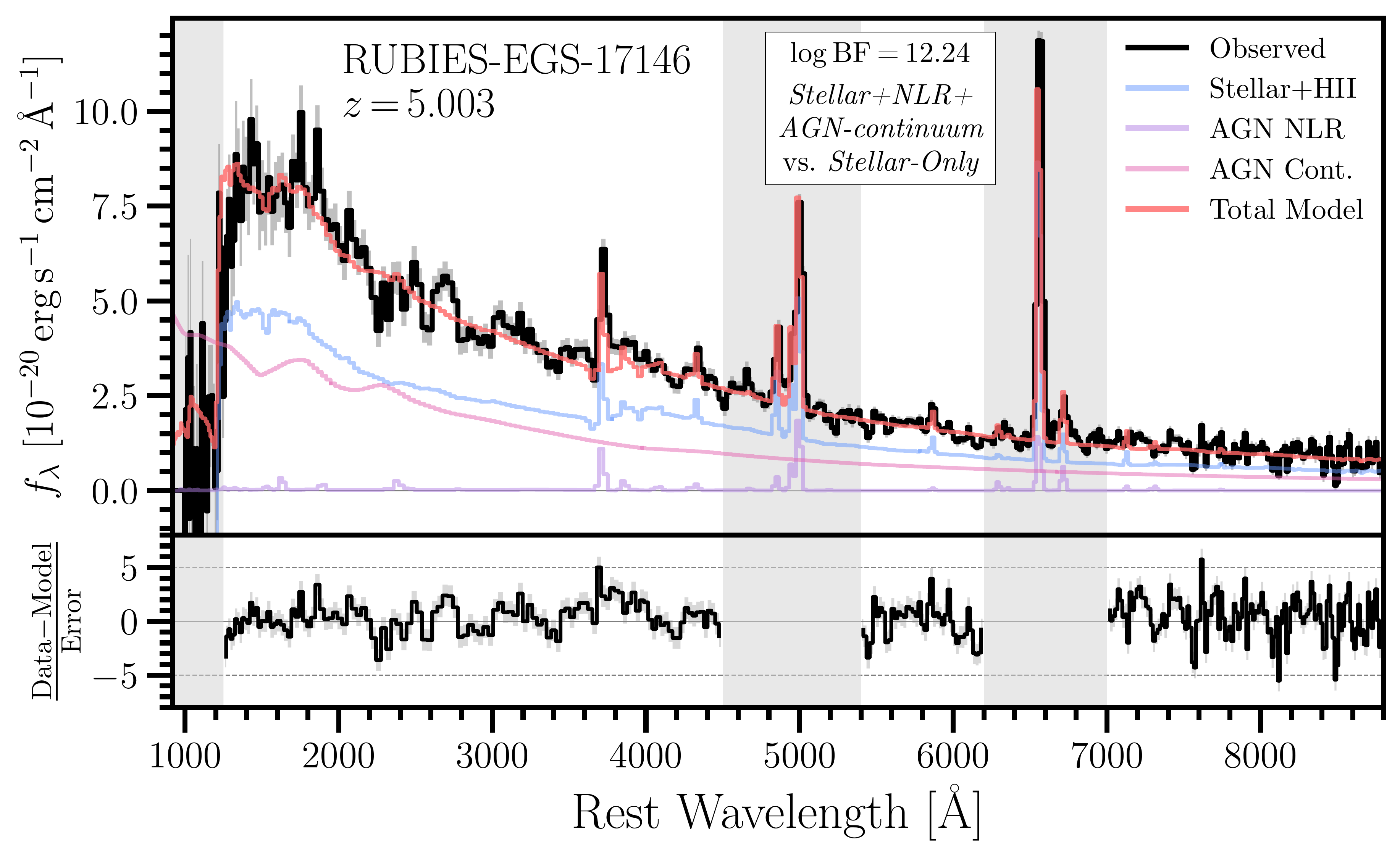}
    \includegraphics[width=\linewidth]{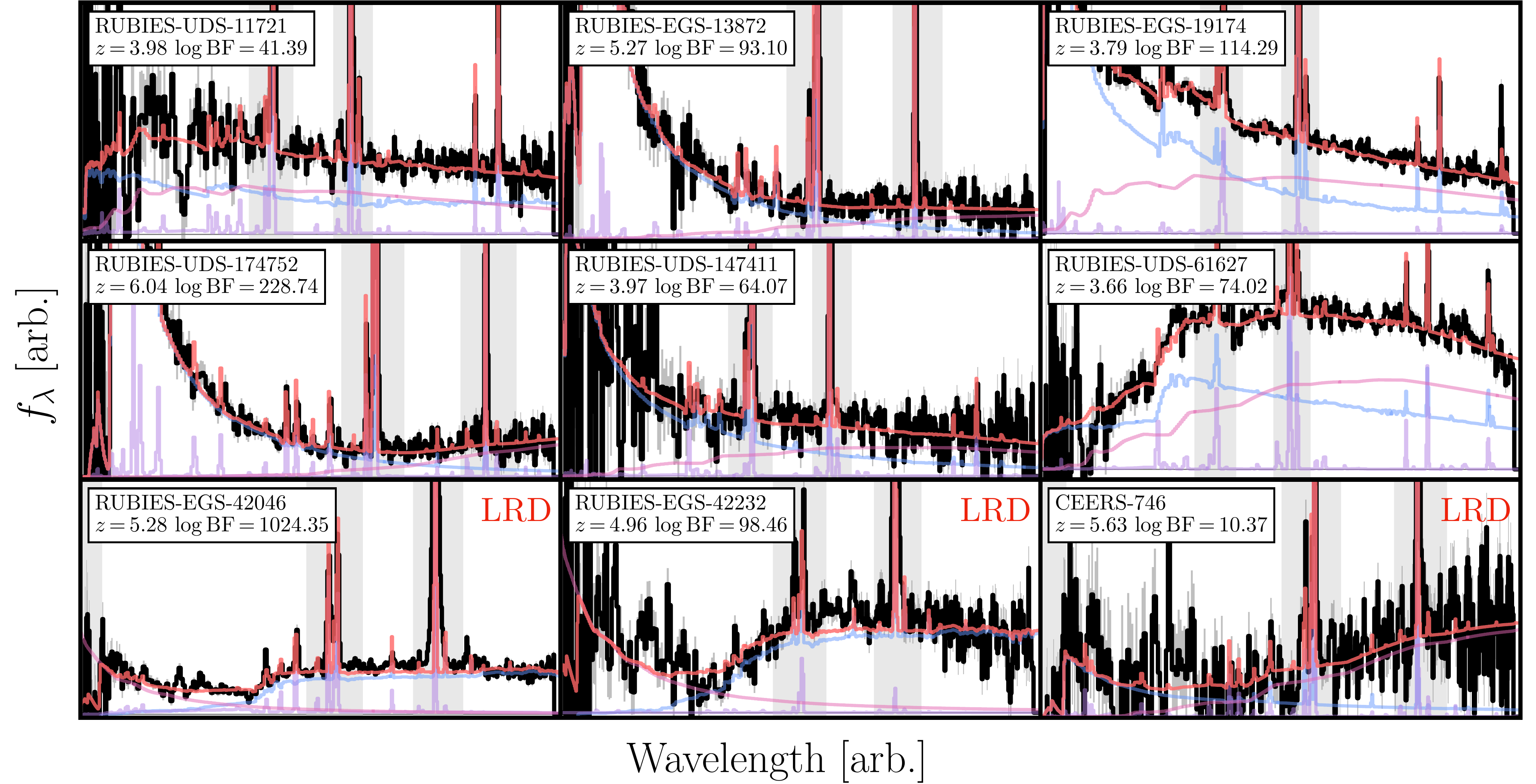}
    \caption{\textit{Top}: An example \beagle\ fit to a BL AGN from our sample, RUBIES-EGS-17146. The observed spectrum is shown in black, with masked portions indicated by shaded regions. The best-fit model (red) includes emission from stars and \HII\ regions (blue) and an AGN component consisting of NLR emission (purple) and a dust-attenuated power-law continuum (magenta). Our best fit reproduces the continuum emission and observed emission lines well, with a significantly better goodness of fit than \textit{stellar-only} and \textit{stellar+NLR} models as measured by a BF comparison. \textit{Bottom}: A variety of \beagle\ fits to our sample. Our modeling suggests that many of these sources are composites, exhibiting both significant stellar-driven and AGN emission. The top two rows show non-LRD AGN sources, while the bottom row displays LRDs. Overall, these results suggest that our modeling is flexible enough to reproduce a wide variety of high-redshift AGN spectra.}
    \label{fig:example-beagle-fit}
\end{figure*}

\subsection{Comparison of Models}\label{sec:4.1}
The \textit{Stellar+NLR+AGN-continuum} models are most preferred for 36 out of 39 of our BL AGN. The \textit{Stellar+NLR} model is preferred for the three remaining sources. Unsurprisingly, these results suggest a strong preference for AGN emission in the best-fit SED models of our sample of BL AGN. Example fits to several BL AGN in our sample using the \textit{stellar+NLR+AGN-continuum} model are shown in Figure \ref{fig:example-beagle-fit}.

Interestingly, we find that the \textit{stellar+NLR+AGN-continuum} models are selected as the best fit for 13 out of our 23 control galaxies, and the \textit{stellar+NLR} models provide the best fit for an additional five control galaxies. The \textit{stellar-only} models are only selected as the best fit for five control galaxies. We discuss potential explanations for this in Section \ref{sec:5.3}.

\begin{figure}[t!]
    \centering
    \includegraphics[width=\linewidth]{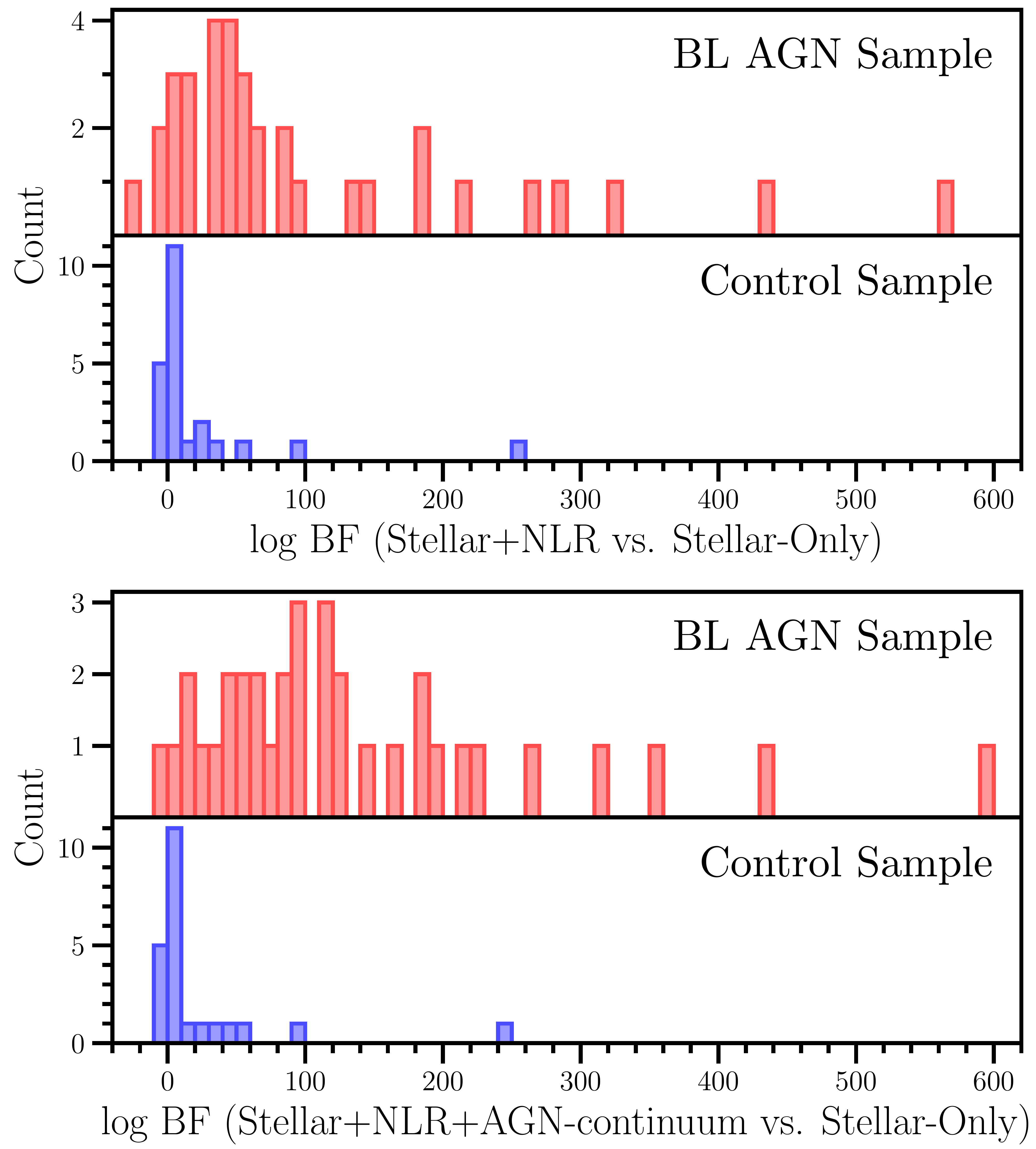}
    \caption{\textit{Top:} Comparison of BF measurements between the \textit{stellar+NLR} model and \textit{stellar-only} model for the BL AGN sample and the control sample. \textit{Bottom:} Comparison of BF measurements between the \textit{stellar+NLR+AGN-continuum} model and \textit{stellar-only} model for the BL AGN sample and the control sample. In both cases, the BF distributions demonstrate that the statistical preference for the AGN models is significantly stronger for the BL AGN sample than for the control sample.}
    \label{fig:bf-comparison}
\end{figure}

\begin{figure}[t!]
    \centering
    \includegraphics[width=\linewidth]{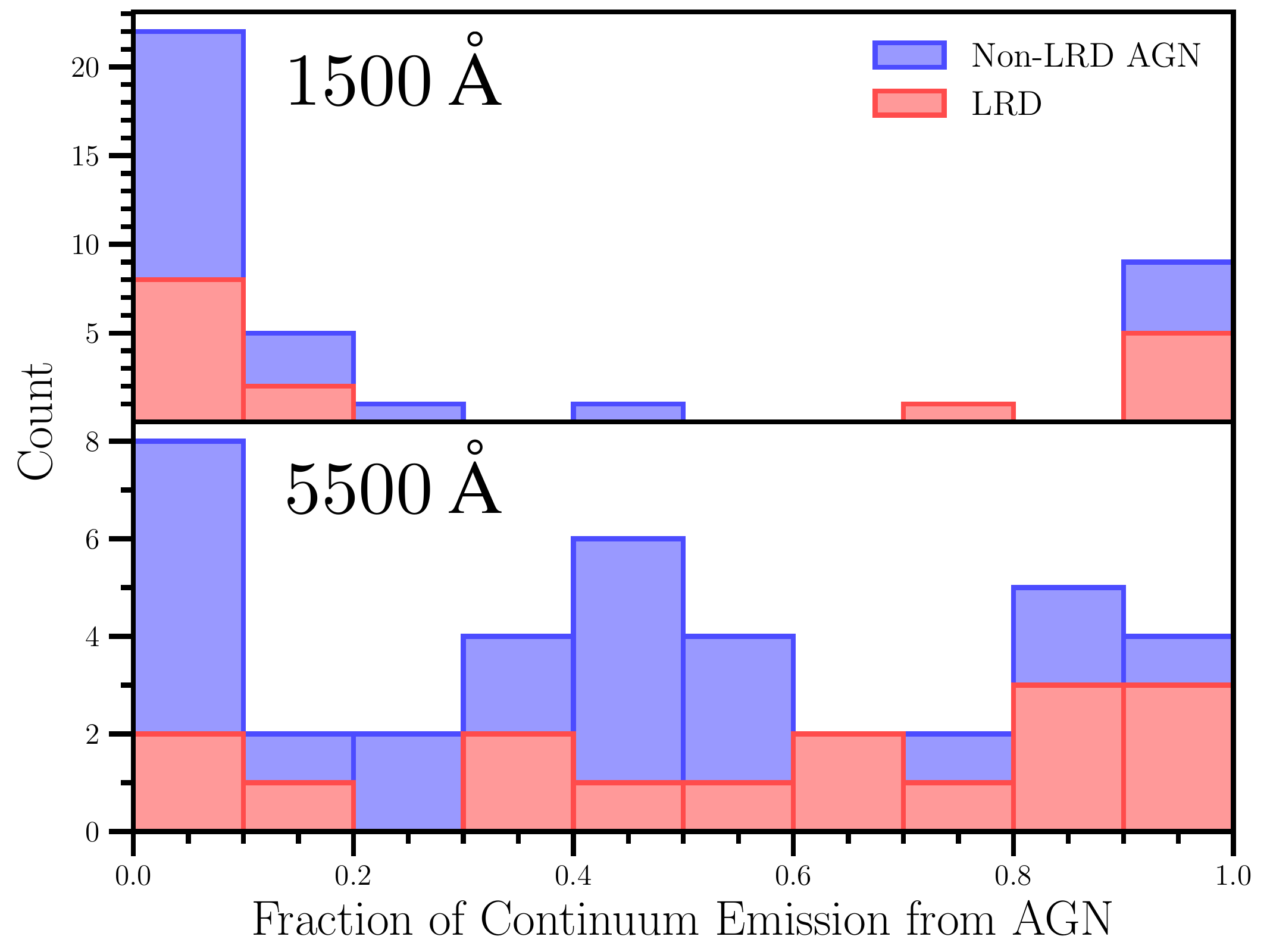}
    \caption{Histograms showing the distribution of the fraction of continuum luminosity from AGN emission in the best-fit \textit{stellar+NLR+AGN-continuum} models in our BL AGN sample, broken up into LRDs (red shaded bars) and non-LRD AGN (blue shaded bars). \textit{Top:} Continuum luminosity at 1500\,\AA. The AGN emission is generally subdominant in the rest-UV, though a moderate fraction of LRDs have a dominant AGN contribution. \textit{Bottom:} Continuum luminosity at 5500\,\AA. Among the majority of both LRDs and non-LRD BL AGN, the rest-optical AGN continuum emission is at least significant, and in many objects it is dominant.}
    \label{fig:continuum-fraction}
\end{figure}

Regardless of the cause of this trend, we derive an important result: for our comparison sample, the \textit{stellar+NLR} and \textit{stellar+NLR+AGN-continuum} models are preferred over the \textit{stellar-only} models by a median log(BF)\,$\sim$\,5. On the other hand, for our main BL AGN sample, the \textit{stellar+NLR} model is preferred by a median log(BF)\,$\sim$\,50 and the \textit{stellar+NLR+AGN-continuum} models by a median log(BF)\,$\sim$\,100 (Figure \ref{fig:bf-comparison}). These findings demonstrate that, although models with AGN emission are slightly preferred for many galaxies in the control sample, the significance of this statistical preference is far greater for the BL AGN sample. This provides strong evidence that our use of NLR and AGN continuum emission to model BL AGN spectra is appropriate and physically motivated, and their preference is unlikely to be due to overfitting. Moreover, even when AGN emission is invoked in the best-fit models for control galaxies, it is generally sub-dominant compared to the stellar emission, a trend that is nearly reversed among the BL AGN sample. For example, examining the \textit{stellar+NLR+AGN-continuum} models, we find that the fraction of optical continuum luminosity at 5500\,\AA\ from the AGN component is at least 30\% in only three out of 23 control galaxies ($\sim$\,13\%), while this threshold is exceeded in 27 out of the 39 sources in the BL AGN sample ($\sim$\,69\%; Figure \ref{fig:continuum-fraction}).

\subsection{\mstar\ Estimates}\label{sec:4.2}
\begin{figure*}[t!]
    \centering
    \includegraphics[width=\linewidth]{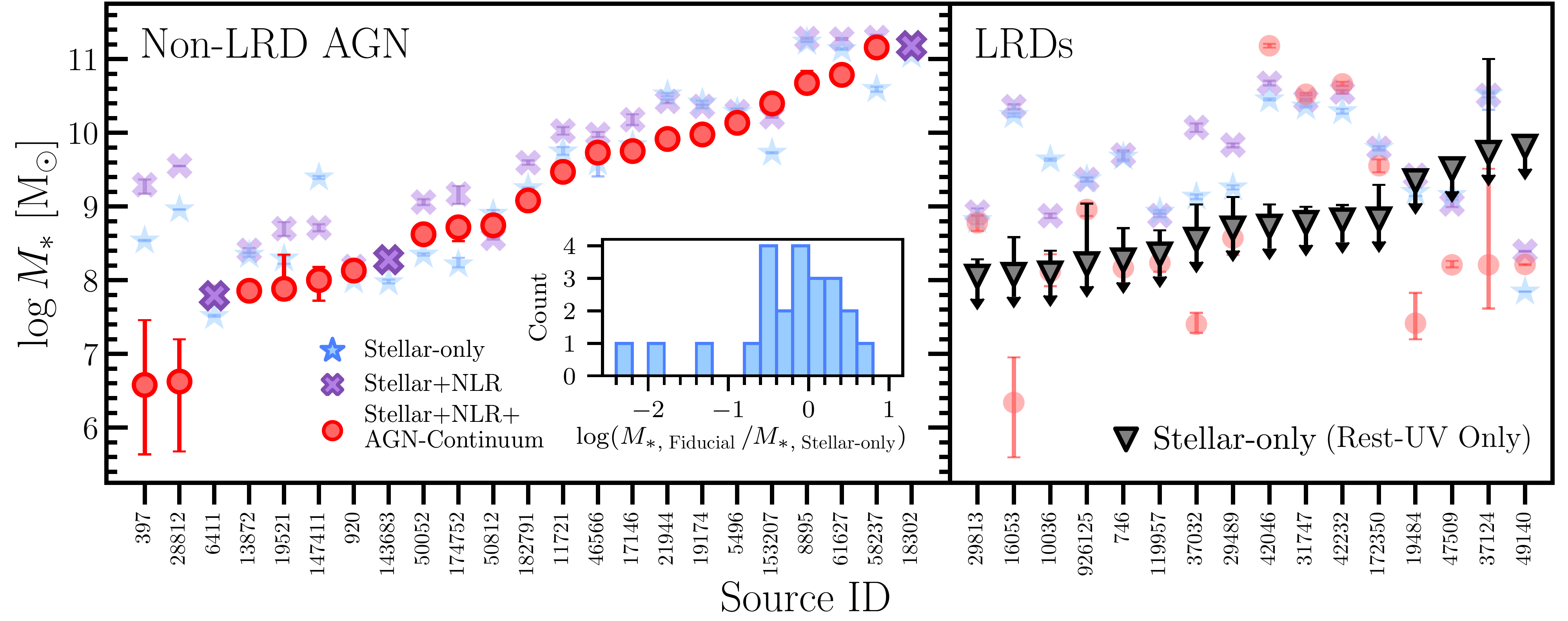}
    \caption{A visual representation of the \mstar\ estimates for fiducial models (full opacity) and other fits (semi-transparent) for each source in our sample, categorized by non-LRD AGN vs.\ LRDs and sorted in order of increasing \mstar\ of the fiducial model. For all but three non-LRD AGN, the \textit{stellar+NLR+AGN-continuum} model provides the best fit. For LRDs, given uncertainties in the nature of the rest-optical emission, we instead measure an upper limit on \mstar\ by performing a \textit{stellar-only} fit to only the rest-UV spectrum (assuming this originates from a host galaxy, with contributions from AGN or other light making this an upper limit; see Section \ref{sec:4.2.2}) and define this as the fiducial model. The differences in \mstar\ estimates between the three models for non-LRDs are modest, with a dispersion of $\sim$\,0.75\,dex. However, the model that includes an AGN continuum does generally yield a lower estimate of \mstar\ than the others. The inset in the left panel shows a histogram of the fiducial \mstar\ value compared to the value derived from the \textit{stellar-only} fit to each non-LRD, visually demonstrating this fact. On the other hand, \mstar\ estimates in LRDs strongly differ between runs which include the AGN continuum and those that do not, which may reflect the inability of publicly-available SED codes to model their underlying physical conditions.}
    \label{fig:mstar-comparison}
\end{figure*}

Now that we have compared the goodness of our model fits, we next examine their derived host stellar masses. For some sources, the \mstar\ estimates derived from each of the three models are roughly consistent, while for others there are significant discrepancies. We first investigate the drivers of differences in \mstar\ estimates between models in non-LRDs, and thereafter examine the LRD population. A visual comparison of \mstar\ estimates between the fiducial model for each source and the alternative fits is shown in Figure \ref{fig:mstar-comparison}.

\subsubsection{\mstar\ Estimates of Non-LRDs}
We first restrict the analysis to the 23 BL AGN that are not LRDs. The differences in \mstar\ estimates between the \textit{stellar-only} and \textit{stellar+NLR} runs are less than 1\,dex for all non-LRDs, and the majority differ by 0.5\,dex or less. One might naively expect that including emission from an AGN would result in a lower \mstar\ estimate. Instead, we observe that 19 out of 23 ($\gtrsim$80\%) of our \textit{stellar+NLR} models yield a \mstar\ value that is \textit{greater} than the corresponding \textit{stellar-only} model. There are a few potential causes of this behavior, but the most significant factor appears to be that the \textit{stellar+NLR} fits infer significantly more dust attenuation than the corresponding \textit{stellar-only} models, with a median A$_{\rm V}$\,$\sim$\,1\,dex higher. This appears to be primarily driven by the presence of NLR emission, as the AGN accretion disk luminosity in the \textit{stellar+NLR} models is found to strongly correlate with A$_{\rm V}$ (as well as with the difference in A$_{\rm V}$ between the \textit{stellar+NLR} and \textit{stellar-only} models). This can be understood in the following way: the NLR emission models have intrinsically bright UV and optical emission lines. However, while the best-fit \textit{stellar+NLR} models of our sources generally invoke AGN NLR emission as a dominant contribution to the observed bright rest-optical emission lines, \jwst-observed high-redshift AGN, including those in our sample, display relatively weak high-ionization UV emission lines \citep[e.g.][]{lambrides_case_2024, wang_missing_2025, zucchi_black_2026}. Dust is thus applied to reconcile differences in observed emission line strengths between the UV and optical regimes. Since we fix the attenuation experienced by the NLR emission to that from stars (see Section \ref{sec:5.2}), the greater extinction in the \textit{stellar+NLR} model means that more intrinsic stellar emission is required to explain the continuum, which increases the inferred \mstar\ values.

We next examine \mstar\ estimates from the \textit{stellar+NLR+AGN-continuum} models, which are selected as the fiducial model for 20 out of 23 ($\gtrsim$85\%) non-LRDs. These models mostly prefer a modestly dust-attenuated (A$_{\rm V}$\,$\sim$\,1--3) AGN continuum, which generally provides a significant contribution across the entire spectrum but tends to be particularly dominant in the rest-optical (Figure \ref{fig:continuum-fraction}). This leaves stellar emission mostly responsible for the rest-UV emission, though with a modest or even dominant rest-optical contribution in some cases. For the 20 non-LRDs in which the \textit{stellar+NLR+AGN-continuum} fit is selected as the fiducial model, 14 (70\%) have \mstar\ estimates lower than those from the corresponding \textit{stellar-only} model (and \textit{stellar-only} \mstar\ estimates for three additional sources are within 0.3\,dex). This figure is even more striking when comparing to the \textit{stellar+NLR} models; \mstar\ estimates in sources with the AGN continuum run as the fiducial model are lower than 18 out of the 20 (90\%) \mstar\ estimates from the corresponding \textit{stellar+NLR} fits (the other two sources have \mstar\ estimates within 0.2\,dex). Overall, it appears that the lower stellar masses derived in the \textit{stellar+NLR+AGN-continuum} models compared to the \textit{stellar-only} models are primarily caused by the AGN continuum displacing stellar emission, as we observe a tight correlation between the fraction of the 5500\,\AA\ continuum luminosity from AGN emission and the difference in stellar mass between the \textit{stellar-only} and the \textit{stellar+NLR+AGN-continuum} models. As opposed to the \textit{stellar+NLR} models, the \textit{stellar+NLR+AGN-continuum} models are only modestly more dusty than the corresponding \textit{stellar-only} models, with a median A$_{\rm V}$\,$\sim$\,0.36\,dex higher. The differences in \mstar\ estimates also do not appear to be driven by a change in the inferred SFH, as the mass-weighted ages of the \textit{stellar-only} and \textit{stellar+NLR+AGN-continuum} models have a median difference within $\sim$\,10\%. Thus, it appears that displacement of the stellar continuum by AGN emission is primarily responsible for the decreased \mstar\ values found in the \textit{stellar+NLR+AGN-continuum} models.

\subsubsection{\mstar\ Estimates of LRDs}\label{sec:4.2.2}

For the LRDs in our sample, the fits which include an AGN continuum component (modeled by a dust-attenuated power-law) result in a significantly improved best-fit model compared to those which attribute the entire continuum to stellar and \HII\ region emission. This result is in agreement with evidence for emission in LRDs being composed of multiple physically distinct components (as described in Section \ref{sec:1}). Previous efforts have failed to describe the continuum shape of LRDs with stars alone, even allowing extremely flexible SFHs and nearly-unphysical dust laws \citep[e.g.][]{ma_uncover_2025}. We find a similar result -- though we allow for multiple populations of stars to be fitted simultaneously (albeit with a somewhat limited parametric SFH prescription), our \textit{stellar-only} and \textit{stellar+NLR} fits are unable to effectively reproduce the continuum shapes of these LRDs.

Our \textit{stellar+NLR+AGN-continuum} fits are more effective at modeling the observed LRD spectra. They are selected as the fiducial model by our Bayes factor comparison test for all LRDs. In general, we observe that the best-fit models attribute most of the rest-optical emission to a modestly dust-attenuated (A$_{\rm V}$\,$\sim$\,1.5--2.5) AGN continuum. For sources with a strong Balmer break feature, the model prefers to include a substantial evolved stellar population (with stellar age above $\sim$\,100\,Myr). In all cases, the rest-UV is reproduced well by a relatively dust-free (A$_{\rm V}$\,$<$1) young stellar population. Though this physical picture superficially matches two-component interpretations of LRD spectra \citep[e.g.][]{degraaff_little_2025, sun_little_2026, baggen_connecting_2026}, many of our fits remain unsatisfactory. Specifically, \beagle\ is generally unable to match the Balmer break feature in sources which display it, even when including a significant evolved stellar population. This is relatively unsurprising, given that the shape of the break in LRDs is smoother than a prototypical stellar Balmer break, and they can also be far stronger \citep[e.g.][]{ji_blackthunder_2025, naidu_black_2025, degraaff_remarkable_2025, taylor_caperslrdz9_2025}. The break feature in LRDs is commonly modeled as AGN continuum emission reprocessed through extremely dense ($n_{\rm H}$\,$>$\,$10^{8}$\,cm$^{-3}$) gas \citep[e.g.][]{inayoshi_extremely_2025} or as blackbody emission from a thermalized ``atmosphere'' \citep[e.g.][]{kido_black_2025}. Current implementations of public SED fitting codes are not generally capable of reproducing such conditions. Previous efforts to model these scenarios by modifying existing SED fitting codes have been successfully deployed \citep[e.g.][]{taylor_caperslrdz9_2025}, but these highly-tailored approaches have not been thoroughly tested or validated for use on broader samples. Although such analysis is outside the scope of this work, we highlight it as an important area for future investigation. In general, our results provide further evidence that, even with sophisticated and flexible modeling techniques, severe caution is warranted in interpreting the \mstar\ estimates of LRDs.

Given these major uncertainties in understanding LRD SEDs (and their underlying physical nature), we do not select any of the \mstar\ values from our primary models as their fiducial measurement. As mentioned previously, high-redshift AGN observed by \jwst\ (especially LRDs) have highly constrained non-detections of high-ionization lines typically associated with AGN \citep[e.g.][]{lambrides_case_2024, wang_missing_2025, zucchi_black_2026}, which is consistent with a stellar origin for the rest-UV. Thus, we measure an upper limit on \mstar\ by performing a \beagle\ \textit{stellar-only} fit constrained to only the spectrum blueward of the Balmer limit at 3645\,\AA. If some fraction of the UV continuum emission is actually non-stellar, then our \mstar\ estimates (which assume a purely stellar origin) will be too high. Thus, we consider these measurements to be upper limits. These upper limit \mstar\ measurements are compared to the derived values from our other runs in Figure \ref{fig:mstar-comparison}.

\begin{figure*}[t!]
    \centering
    \includegraphics[width=\linewidth]{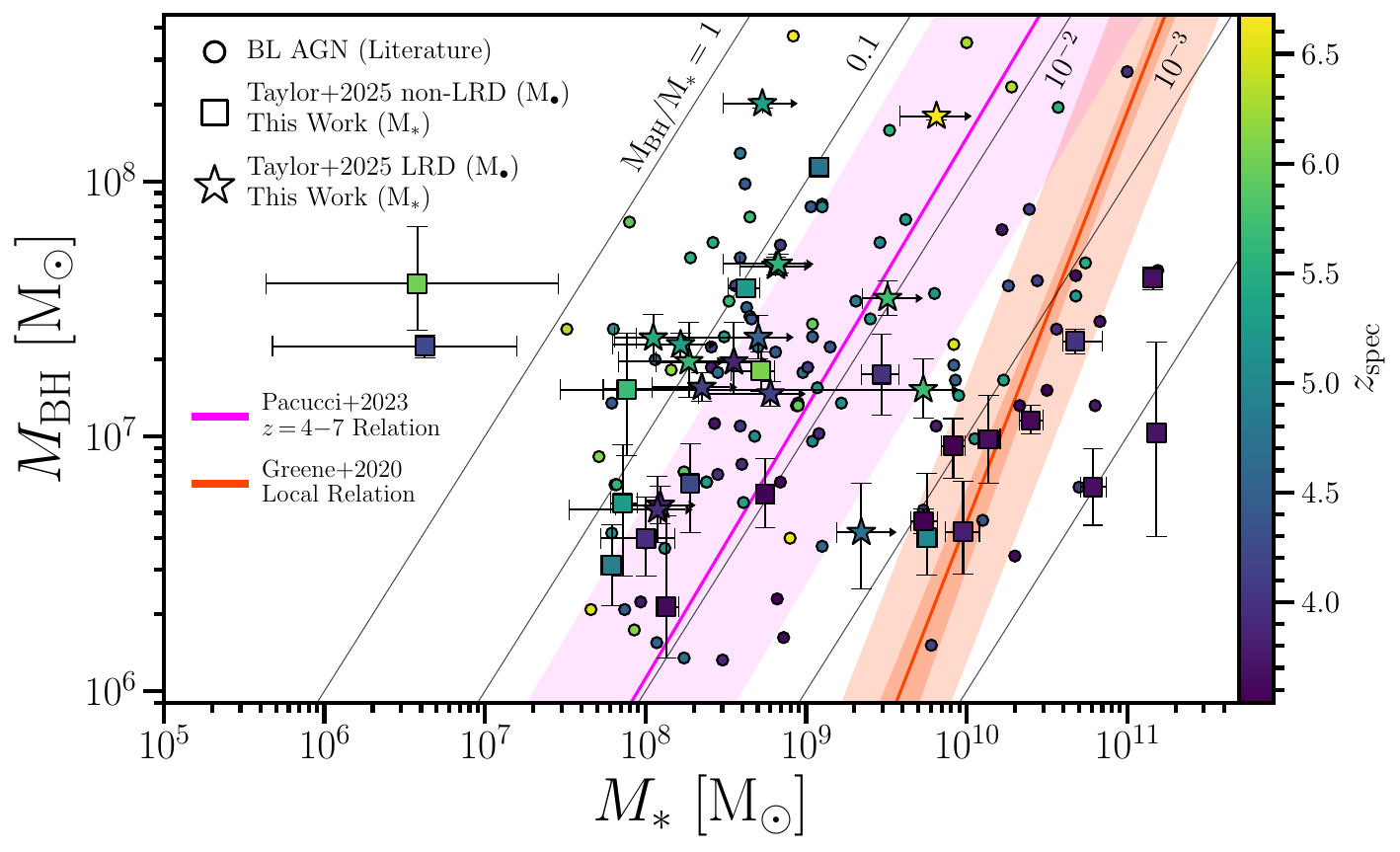}
    \caption{The \mbhmstar\ relation for AGN observed by \jwst. Sources analyzed in this work are shown using square markers (non-LRDs) or stars (LRDs), while objects reported in previous studies (Appendix \ref{apdx:c}) within the redshift range of our sample (3.5\,$\lesssim$\,$z$\,$\lesssim$\,7) are drawn as smaller circles. Sources are color-coded by spectroscopic redshift. There appears to be an approximately bimodal distribution of \mbhmstar\ values, with low-\z\ sources clustered near the local relation \citep[orange;][]{greene_intermediatemass_2020} and high-redshift sources displaying elevated \mbhmstar\ ratios in line with the empirical high-redshift best-fit \citep[magenta, \z\,$\sim$\,4--7;][]{pacucci_jwst_2023}. This redshift dependence is explored further in Figure~\ref{fig:mbhmstar-z}.}
    \label{fig:mbh-mstar}
\end{figure*}

\begin{figure*}[t!]
    \centering
    \includegraphics[width=\linewidth]{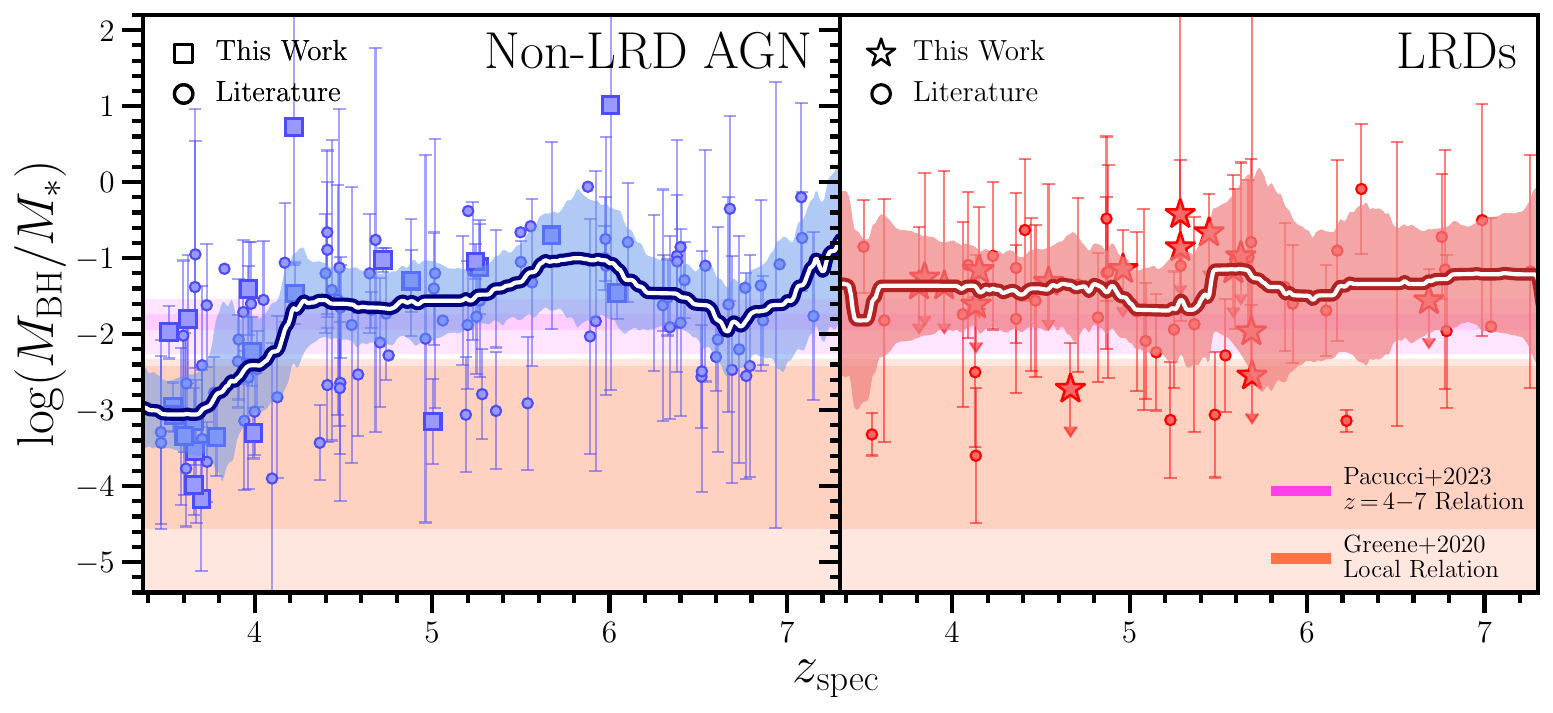}
    \caption{The \mbhmstar\ ratio plotted against spectroscopic redshift for our sample (squares for non-LRDs and stars for LRDs) and literature BL AGN (circles, Appendix \ref{apdx:c}), divided into non-LRD AGN and LRDs. Regions consistent with sources following the local relation \citep[pale orange shading;][]{greene_intermediatemass_2020} and high-redshift relation \citep[pink shading;][]{pacucci_jwst_2023} are also shown. A smoothed continuous median for each set of data derived from ``pseudo-binning'' is overplotted (see Section \ref{results:mbh-mstar} for details). The \mbhmstar\ ratio of non-LRD AGN appears to evolve from an value elevated by $\sim$\,1\,dex at \z\,$\gtrsim$\,$4.5$ to one consistent with the local relation at \z\,$\lesssim$\,$3.5$ (a period of $\sim$\,500\,Myr). Due to the complexity in characterizing observational completeness across this heterogeneous sample, a full exploration of systematics is outside the scope of this work. As such, while we cannot entirely rule out selection effects as the cause of this apparent evolution, the short timescale of this transition means that any systematic effects must vary rapidly as a function of redshift. In contrast, the \mbhmstar\ relation of LRDs appears to remain remarkably flat across time, which may reflect limitations in our ability to estimate their \mbh\ (if their emission is indeed driven by accreting BHs) and/or \mstar\ given their uncertain physical nature.}
    \label{fig:mbhmstar-z}
\end{figure*}

\subsection{The high-redshift \mbhmstar\ Relation}\label{results:mbh-mstar}
We next compare the \mstar\ value of each source (from its fiducial model) with \mbh\ estimates from \citet{taylor_broadline_2025}. \jwst-detected BL AGN from our sample and from the literature are plotted on the \mbhmstar\ relation in Figure \ref{fig:mbh-mstar}. Taking the \mbh\ estimates at face value, a substantial fraction of AGN in our sample appear to host BHs with masses $\sim$\,1--2\,dex higher than expected from the local relation given their stellar masses. Sources appear to generally lie along two loci in the space of \mbh\ vs.\ \mstar, with one cluster near the local relation and another at an elevated ratio of \mbhmstar\,$\sim$\,0.05--0.1. The set of non-LRDs with overmassive BHs are heavily skewed towards the high-redshift end of our sample, whereas low-\z\ non-LRDs nearly all exist near the local relation. In contrast, LRDs nearly all have BHs which are overmassive by $\gtrsim$\,1\,dex, regardless of redshift. One significant outlier is RUBIES-EGS-28812, which \citet{taylor_broadline_2025} report as a non-LRD at a relatively low redshift of \z\,=\,4.22, but which has an estimated \mbhmstar\ value in excess of unity. However, this source is selected as an LRD using the slightly looser criteria of \citet{barro_extremely_2024}, and displays a rest-optical continuum that is well-described by a modified blackbody spectrum. If this source is indeed an LRD, this would negate the significance of its high \mbhmstar\ ratio at such a low redshift. Overall, these results are consistent with findings from the existing literature \citep[e.g.][]{pacucci_jwst_2023, jones_m_rm_2025, juodzbalis_jades_2026}; this is somewhat unsurprising, given that our \mstar\ measurements are roughly consistent with those from previous studies, but the expanded sample size afforded by our analysis helps clarify this.

To better quantify the observed trends, we show the \mbhmstar\ ratio of our sources and \jwst-observed AGN from the literature (see Appendix \ref{apdx:c}) as a function of redshift in Figure \ref{fig:mbhmstar-z}, separated into LRDs and non-LRD AGN. To quantify any evolution in the median value of \mbhmstar\ as a function of redshift, we pool our sample with a comprehensive sample of published \jwst-observed AGN and derive a smooth binned median using the ``pseudo-binning'' technique \citep[e.g.][]{finkelstein_census_2022}. This process approximates the results of a bootstrapped binned median, but marginalizes over choices of bin width and position. For each point in a finely spaced grid of redshift values (d\z\,=0.01), we randomly draw $N$\,=\,1000 sample bin widths ranging from $\Delta z$\,=\,0.1--1. The median value of \mbhmstar\ for points that fall in each sample bin is computed, and aggregated together these form a distribution of median \mbhmstar\ values at each point in redshift. Finally, the 50th percentile of this distribution is used to build a smooth curve representing the bootstrapped median across the redshift range, with the 16th and 84th percentile values constituting the 1\,$\sigma$ confidence interval of this quantity. We add Poisson errors for each bin and the measurement uncertainties of the sources in each bin in quadrature with this confidence interval to form a 1\,$\sigma$ error, and we finally apply Gaussian smoothing with standard deviation $\sigma_z$\,=\,0.01 for plotting purposes. The bootstrapped median curves along with the 1\,$\sigma$ error regions are shown as the solid lines and shaded regions in Figure \ref{fig:mbhmstar-z}.

We emphasize that the plotted dispersion is \emph{not} our claimed uncertainty in the \mbhmstar\ relation. Instead, this measurement quantifies our ability to estimate the median of the observed distribution.

Some qualitative trends can immediately be observed. Among non-LRD AGN, we find an evolution in observed \mbhmstar\ as a function of redshift. Specifically, observed \mbhmstar\ ratios generally appear elevated and relatively flat at \z\,$\gtrsim$\,$4.5$, but rapidly evolve towards a value consistent with the local relation by \z\,$\sim$\,$3.5$. On the other hand, the \mbhmstar\ relation for LRDs appears to be nearly flat as a function of redshift. While we can directly observe these trends in the \textit{observed} \mbhmstar\ relation, we discuss implications for the \textit{intrinsic} high-redshift \mbhmstar\ relation in Section \ref{discussion:mbh-mstar}.

\section{Discussion}\label{sec:5}

\subsection{Evolution of the \mbhmstar\ Relation}\label{discussion:mbh-mstar}
In Section \ref{results:mbh-mstar}, we describe the implications of our newly-derived \mstar\ estimates on the \textit{observed} \mbhmstar\ relation, noting that it appears flat at \z\,$\gtrsim$\,4.5 with a rapid transition to the local relation between \z\,$\sim$\,3.5--4.5. However, due to incompleteness, observational biases, and systematic errors, it is not straightforward to infer the \textit{intrinsic} high-redshift \mbhmstar\ relation from the observed one. For instance, the effect known as the ``Lauer bias'' \citep[][]{lauer_selection_2007} means that AGN-selected samples with intrinsic scatter in \mbhmstar\ will tend to select overmassive BHs, especially at \mstar\ values in steep portions of the galaxy stellar mass function, as there are more low-mass galaxies which can be up-scattered in \mbhmstar\ than high-mass galaxies which are available to be down-scattered. Moreover, estimating the selection function of our sample (especially pooled together with the total literature sample) is highly nontrivial. One major challenge lies in estimating the bias introduced by the choice of NIRSpec  micro-shutter assembly (MSA) mask positions. Masks are designed to maximize the number of high-priority targets on slits, which might introduce selection effects. For instance, if AGN hosting overmassive BHs appear redder and more luminous, then they may be up-weighted in mask selection. MSA positions are influenced by all targets (not just AGN) across a NIRSpec pointing, making accounting for this factor extremely challenging.

\citet{taylor_broadline_2025} estimate the completeness of their sample by accounting for two factors: ``line detection'' incompleteness, which describes when true BL AGN with spectroscopic observations are not selected as such, and ``observational incompleteness'', which results from candidate AGN in the MSA area not being assigned open shutters (and thus failing to be spectroscopically observed). Unfortunately, extending this measure of completeness to vary as a function of \mbhmstar\ is far more complicated, and likely requires a full forward-modeling approach taking account of both physical factors such as the Eddington ratio distribution function and observational biases and potential systematic errors \citep[similar to e.g.][]{pacucci_jwst_2023, li_tip_2025, silverman_shellqsjwst_2025}. As such, we defer a full quantitative investigation of the redshift evolution in the intrinsic \mbhmstar\ relation to a future paper.

Here, we make note of two relevant findings. The first is that we find modest, but significant, differences in \mstar\ estimates between various models. For instance, among non-LRDs, the distribution of the differences in \mstar\ estimates between the \textit{stellar-only} models and the \textit{stellar+NLR+AGN-continuum} models has a standard deviation of $\sim$\,0.75\,dex (see inset in the left panel of Figure \ref{fig:mstar-comparison}). Any errors caused by incorrect model assumptions will also be combined with uncertainties in \mstar\ estimation between various SED fitting codes \citep[e.g.][]{pacifici_art_2023}, something that has been explicitly examined in \mstar\ measurements of \jwst-detected AGN \citep{juodzbalis_jades_2026}. Further discrepancies can be introduced by other methodological differences such as fitting of photometry instead of spectroscopy or by the use of AGN-host galaxy spatial decomposition. Clearly, there may be major systematic uncertainties in \mstar\ measurements. As demonstrated in \citet{li_tip_2025}, an uncertainty in log(\mstar) of $\sim$\,0.45\,dex is sufficient to cause a population of high-redshift AGN drawn from a ``normal'' \mbhmstar\ distribution to appear to display elevated \mbhmstar\ ratios after selection effects and mass measurement uncertainties are applied.

\begin{figure}[t!]
    \centering
    \includegraphics[width=\linewidth]{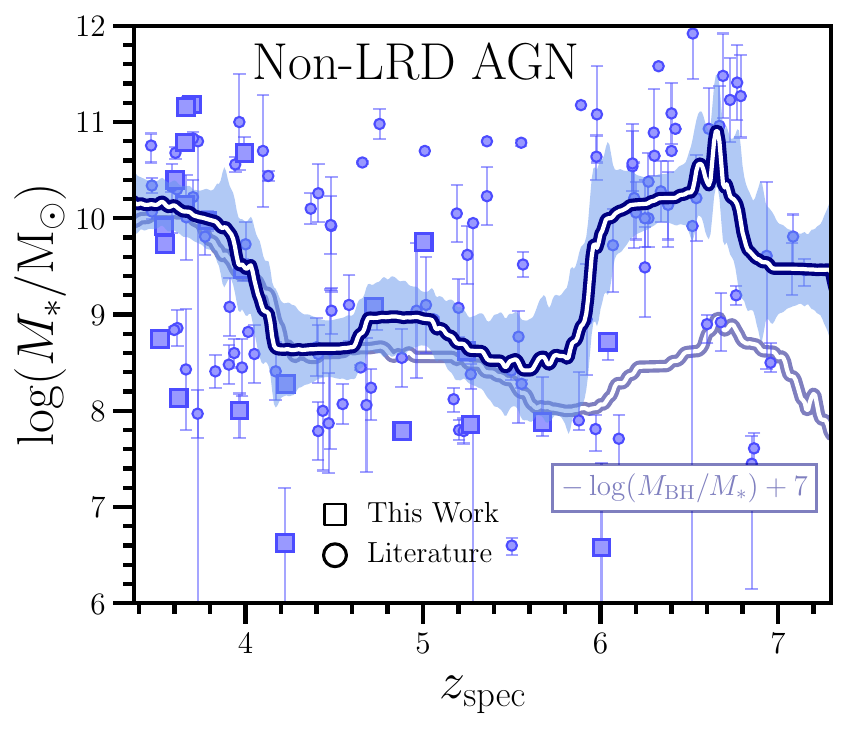}
    \caption{\mstar\ as a function of redshift for non-LRD AGN. Sources analyzed in this work are shown using square markers, while objects reported in previous studies (Appendix \ref{apdx:c}) are drawn as circles. A smooth binned median generated using the pseudo-binning technique (see Section \ref{results:mbh-mstar} for details) is overplotted in solid blue. The faded blue curve shows an inverted and re-normalized version of the smoothed \mbhmstar\ curve for non-LRDs from Figure \ref{fig:mbhmstar-z} for comparison. Though galaxies with $\log (M_{*}$/M$_{\odot})$\,$\lesssim$\,$8$ are observed up to \z\,$\sim$\,6, there is a rapid change in the median \mstar\ value of AGN between \z\,$\sim$\,3.5--4.5. Across this same redshift range, the distribution of inferred \mbh\ of AGN is relatively flat, meaning that this evolution in measured \mstar\ drives the evolution in the observed \mbhmstar\ relation.}
    \label{fig:mstar-z}
\end{figure}

However, we now focus on a second finding: the rapid evolution in the observed \mbhmstar\ relation between \z\,$\sim$\,3.5--4.5. It is unclear what combination of systematic and observational uncertainties would drive such a sharp change, when AGN hosting relatively ``undermassive'' BHs can and have been observed at these redshifts \citep[e.g.][]{li_prevalent_2025, sun_no_2025, zhang_abundant_2026}. Moreover, we find that the evolution in the observed \mbhmstar\ relation is entirely driven by changes in \mstar\ of AGN, as shown in Figure \ref{fig:mstar-z}, while the estimated \mbh\ distribution across the redshift range of \z\,$\sim$\,3.5--4.5 is flat. One might expect that a BH of a given mass and accretion rate embedded in a fainter galaxy may be easier to detect at high-\z, as it will outshine its host more strongly and is more likely to be selected as an AGN via photometric colors and compactness. However, it is unclear why a selection effect like this would result in such a discontinuous shift in the observed \mstar\ distribution of AGN from \z\,$\sim$\,3.5--4.5, while flattening out to higher redshifts. Moreover, if the evolution in the observed \mbhmstar\ relation is indeed caused by selection effects, this would imply that the scatter in the intrinsic high-redshift \mbhmstar\ relation is much larger than is observed in the local universe, with currently observed AGN tracing the upper envelope of the underlying distribution \citep[e.g.][]{sun_m_m_rm_2025, li_tip_2025, silverman_shellqsjwst_2025}. If the normalization of the high-redshift \mbhmstar\ relation is consistent with the local relation, this would imply the presence of a substantial unobserved population of BHs; however, it has been argued \citep[e.g.][]{guia_no_2024, jones_m_rm_2025, roberts_too_2026} that this may be in some tension with the fact that current \jwst-observed AGN already appear to constitute a substantial fraction of high-redshift galaxies and that the high-redshift BH mass function \citep[BHMF; e.g.][]{matthee_little_2024, taylor_broadline_2025, fei_glimpse_2025} from high-redshift AGN is consistent with the BHMF extrapolated from observations of quasars detected in ground-based observations \citep[e.g.][]{he_black_2024}.

On the other hand, if this trend corresponds to a real change in the intrinsic \mbhmstar\ relation, there are some interesting theoretical precedents for a significant transition within this redshift range. For instance, \citet{trinca_episodic_2024} examine the \mbhmstar\ relation of AGN in the Cosmic Archaeology Tool semi-analytical model and find elevated ratios (similar to those seen in \jwst\ AGN) caused by repeated bursts of super-Eddington accretion driven by major merger events. In the simulation, the decrease in merger frequency by \z\,$\sim$\,4 causes SMBH growth to cease and enables a rapid buildup of the host galaxy, returning systems to the local relation within $\sim$\,500\,Myr. \citet{pacucci_redshift_2024} find similar results but attribute the rapid evolution to secular processes, including AGN feedback. Qualitatively similar behavior is also observed in theoretical models which track the formation and growth of heavy BH seeds \citep[e.g.][]{scoggins_how_2023, bhowmick_growth_2024, jeon_little_2026}. The fact that the evolution in the observed \mbhmstar\ relation is entirely driven by a change in measured \mstar\ is thus intriguing, and may reflect a rapid assembly of AGN host galaxies. Nonetheless, as we are unable to account for all observational biases and systematic uncertainties in this work, we are unable to exclude the possibility that at least part of this observed evolution may be due to selection effects.

\subsection{Limitations in \beagle\ Modeling}\label{sec:5.2}
The greatest differences in \mstar\ estimates are between the \textit{stellar+NLR} models and the \textit{stellar+NLR+AGN-continuum} models. Since we analyze BL AGN, which (at least among non-LRDs) likely present a relatively unobscured view of the central engine, this is somewhat expected. However, it is important to note that our AGN continuum model is relatively simplistic, consisting of an analytical power-law attenuated by dust. We briefly explore the impacts of AGN continuum model choices by investigating differences in parameter estimates between models with $\alpha_{\rm cont}$\,=\,$-1.54$ and a second set of models which are identical except that we set $\alpha_{\rm cont}$\,=\,$-2.33$. We find that, among non-LRDs, the distribution of differences in \mstar\ estimates between runs with these two power-law slopes has a median $\sim$\,0\,dex with standard deviation $\sim$\,0.15\,dex. We thus conclude that our \mstar\ estimates are relatively insensitive to our choice of $\alpha_{\rm cont}$. Future investigations will be required to assess the detailed impact of AGN continuum model priors on the robustness of retrieved galaxy properties, but such analysis is outside the scope of this work.

As described in Section \ref{sec:3.3}, the UV-optical slope of the AGN continuum ($\alpha_{\rm cont}$) is, in general, not the same as the slope of the incident ionizing radiation field on the NLR ($\alpha_{\rm NLR}$). To choose a value of $\alpha_{\rm NLR}$, we look to the results of \citet{vidal-garcia_beagleagn_2024}, who fit a comprehensive set of synthetic NL AGN spectra with \beagle\ to test the reliability of fitting and the robustness of retrieved parameters of the NLR and host galaxy. These findings show that allowing $\alpha_{\rm NLR}$ to vary introduces a degeneracy with the AGN accretion disk luminosity $L_{\rm acc}$, resulting in the latter being overestimated. As \citet{vidal-garcia_beagleagn_2024} obtain the best results fixing $\alpha_{\rm NLR}$\,=\,$-1.7$, we choose the same prior. Nonetheless, different choices of $\alpha_{\rm NLR}$ may yield somewhat different results. Similarly, while the AGN covering fraction $C_f$ is fixed to 10\% in the available models, and this quantity is almost certainly degenerate with quantities such as the total accretion disk luminosity, modifying this value may somewhat impact our results. However, one somewhat reassuring fact is that the impact of including NLR emission in the modeling impacts derived \mstar\ estimates only modestly, and its effects are less pronounced than the inclusion of AGN continuum emission (see Section \ref{sec:4.2}).

In these models, emission from the NLR is attenuated by the diffuse dust component in the ISM (matching the attenuation applied to emission from stars with age $>$\,10\,Myr). In principle, however, the NLR could experience additional attenuation if dust conditions near the AGN differ from the broader ISM of the host galaxy \citep[e.g.][]{heard_location_2016, lu_reddening_2019, brooks_here_2025}. To examine this effect, \citet{silcock_characterising_2025} test the addition of an independent attenuation factor for the NLR, $\hat{\tau}_{\rm V,\,NLR}$. This factor is found to be highly unconstrained and the retrieved parameter estimates do not differ significantly (statistically) from the case in which the NLR attenuation is fixed to the stellar attenuation. As a result, we choose not to fit this parameter in our models.

In the above analysis, we assume that our spectra contain sufficient information for \beagle\ to robustly disentangle emission originating in the NLR from that of the host galaxy. In a line-only modeling analysis fitting optical line fluxes, \citet{vidal-garcia_beagleagn_2024} find that NLR emission can be robustly detected in sources with S/N(\Hb)\,$\sim$\,10, a criterion met by over half of the sources in our sample (with all but a small fraction displaying at least SNR(\Hb)\,$>$\,5). However, these authors also demonstrate that a higher SNR may be required for unbiased parameter retrieval, especially for systems where emission lines are dominated by \HII\ regions. Since we analyze known BL AGN, where the broad lines are generally quite strong compared to the narrow emission, it seems likely that most of these systems have narrow line emission dominated by the NLR rather than by \HII\ regions. As a point of comparison, \citet{malkan_narrow_2026} analyze a large sample of low-\z\ AGN and found that $\sim$90\% of narrow line emission in Type 1 Seyferts (BL AGN) arises from the NLR. Regardless, this caveat suggests that the best-fit values of quantities such as metallicities and ionization parameters of the host galaxy and NLR region should be used with caution. However, we observe no strong trend in the \mstar\ estimates between different models (or their differences) as a function of SNR(\Hb).

An additional related limitation is that integrated photometry and spectroscopy extracted from an individual NIRSpec MSA shutter may sample slightly different mixes of the galaxy ISM and NLR emission, especially in spatially extended sources. The optimally extracted 1D spectra may thus not reflect the entire emission from the host galaxy, potentially in a wavelength-dependent way. Narrow-band imaging or integral field unit spectroscopy of these AGN would allow us to measure the spatial extent of line emission and determine how significant this effect is likely to be. As these data currently do not exist for our sample, we are unable to correct for this effect. However, many of our sources are unresolved in NIRCam imaging, and results do not appear to significantly differ between compact and extended sources.

In this work, we use a parametric delayed-exponential SFH with an IMF cutoff of 100\,M$_{\odot}$. Both of these factors may impact our derived stellar masses. While a more flexible non-parametric SFH may yield different \mstar\ estimates, modeling several independently parameterized SFH bins would require fitting a significantly larger number of parameters which would almost certainly impact the convergence of our fits and potentially bias retrieved parameter estimates. On the other hand, increasing the IMF cutoff to e.g. 300\,M$_{\odot}$ may decrease some derived \mstar\ values, as the highest mass stars would likely decrease the mass to light ratio of the overall stellar population, but a prior exploration of varying the IMF cutoff in \beagle\ found minimal impacts compared to the effects of including AGN models \citep{chavezortiz_significant_2025}.

\subsection{Why are Control Galaxies Better Fitted by Models With AGN Emission?}\label{sec:5.3}
In Section \ref{sec:4.1}, we show that 18 out of 23 galaxies in the control sample are better fit by models with AGN emission than \textit{stellar-only} models. The significance of the statistical preference for AGN models among control sample galaxies is much weaker than the corresponding statistical preference among our BL AGN sample (as measured by a comparison of BF distributions, Figure \ref{fig:bf-comparison}). Nonetheless, this may suggest potential limitations inherent in our \textit{stellar-only} models. Here, we discuss factors which may explain the preference for AGN models over \textit{stellar-only} models for these sources.

One possible explanation is that the ISM conditions of high-redshift galaxies appear distinct from their low-\z\ counterparts \citep[which often anchor the priors used in SED fitting; e.g.][]{chevallard_modelling_2016, leja_deriving_2017, carnall_inferring_2018}. For instance, high-redshift chemical abundances may depart from Solar values \citep[e.g.][]{topping_metalpoor_2024, hsiao_first_2025, berg_fleeting_2025, scholtz_jades_2026}. Moreover, the ISM at high-redshift may be characterized by substantially higher electron densities than in most low-\z\ galaxies \citep[e.g.][]{isobe_redshift_2023, abdurrouf_jwst_2024, topping_aurora_2025}, which could strongly impact diagnostic emission line ratios \citep[e.g.][]{hayes_average_2025, martinez_pressure_2025, hsiao_glimpse_2026}. An additional challenge is that the \HII\ regions of high-redshift sources may not be accurately modeled with single-zone ionized regions; in reality, the ISM of galaxies (especially at high-\z) can be highly stratified, with multiple ionization zones leading to more complex nebular emission. \citep[e.g.][]{zeimann_hubble_2015, berg_characterizing_2021, marconi_homerun_2024, shapley_aurora_2025}. In our fits, we assume a solar carbon abundance with (C/O)$_\odot$\,=\,0.44, a nitrogen abundance anchored to the solar value \citep[though rescaled based on metallicity-sensitive secondary nitrogen production mechanisms;][]{gutkin_modelling_2016}, and an ISM \HII\ region hydrogen density $n_{\rm  H}^{\rm HII}$\,=\,100 cm$^{-3}$. Moreover, \beagle\ characterizes emission from \HII\ regions with a single metallicity and ionization parameter. Sources in our control sample may thus be better fit by models with AGN emission because these modeling choices may not accurately reflect the ISM conditions of high-redshift galaxies. Using extended nebular emission grids, one can explore the impacts of varying chemical abundances \citep[e.g.][]{chavezortiz_significant_2025} and \HII\ region density using \beagle. However, since only the strongest emission lines are significantly detected in the available spectra for our control sample, carrying out this procedure with the currently available data would almost certainly introduce significant degeneracies in the fitting and bias the resulting parameter estimates. While such an exploration is outside the scope of this analysis, future work should be carried out to better understand the impacts of these modeling choices.

There is a second factor which may explain the preference for AGN models in a subset of the control sample. Our selection criteria exclude only objects with detected broad lines; thus, a non-negligible fraction of these sources may be NL AGN \citep{chisholm_ne_2024, scholtz_jades_2025, mazzolari_narrow_2025} and thus naturally better fit by a model which includes AGN emission. Removing NL AGN from the control sample would require applying cuts based on emission line fluxes or diagnostic ratios. However, these are the same factors implicitly used in the \beagle\ fitting to determine the relative contributions of \HII\ regions and NLR emission. Since these galaxies are used as a control sample, it would be somewhat circular and potentially problematic to remove any but the most unambiguous NL AGN. As discussed above, the ISM conditions in high-redshift galaxies are quite different than those at low-\z, meaning that diagnostic diagrams used to select AGN in the local universe lose their constraining power at earlier times \citep[e.g.][]{hirschmann_emissionline_2023, cameron_jades_2023, gupta_emissionline_2024, backhaus_emissionline_2025, cleri_optical_2025}. Some inactive galaxies, especially sources with extreme ISM conditions (e.g. those in the upper envelope of the $\log$\,$U^{\rm HII}$ distribution) may thus be accidentally removed during a NL AGN cut to the control sample. This would artificially inflate the significance of results in our BL AGN sample.

\section{Conclusions}\label{sec:6}
In this work, we examine the evolution of the \mbhmstar\ relation across cosmic time. Previous works have identified many high-redshift AGN with \mbhmstar\ ratios which appear elevated compared to local relations. A major systematic uncertainty impacting these results is the accuracy of \mstar\ measurements. Here, we test the robustness of \mstar\ estimates derived from SED fitting by studying a sample of 39 BL AGN from the CEERS and RUBIES surveys with \jwst/NIRSpec spectroscopic data from both the PRISM and G395M dispersers.

We perform SED fitting directly on PRISM spectra, with additional constraints from the G395M data. We use the \beagle\ SED fitting code to self-consistently model emission from both the host galaxy and the NLR. This allows us to incorporate NLR emission in SED fitting at high-redshift in a large sample for the first time. We find that including models of NLR emission results in an improved best-fit model for all sources. Moreover, including an AGN continuum component modeled by an independently attenuated power-law produces a further improved fit for 36 out of 39 sources. We note the following key results:

\interfootnotelinepenalty=0
\interlinepenalty=0
\begin{itemize}
    \setlength{\leftskip}{1em}
    \setlength{\itemsep}{0pt}
    \item Among non-LRD AGN, models which include NLR emission produce better fits and yield \textit{increased} \mstar\ estimates compared to \textit{stellar-only} models (median \mstar\,$\sim$\,0.24\,dex higher with a dispersion of $\sim$\,0.35\,dex). This is predominantly caused by changes in the inferred dust attenuation applied to the stellar emission.
    
    \item For nearly all non-LRD AGN, the inclusion of an AGN continuum model results in an improved fit, and \mstar\ estimates are reduced (median \mstar\,$\sim$\,0.33\,dex lower with a dispersion of $\sim$\,0.75\,dex) primarily due to a displacement of continuum emission which would otherwise be reproduced by stars. 
    
    \item Our tested models are unable to reproduce the strong break features in some LRD spectra, a result in qualitative agreement with results from previous works fitting them using public SED fitting codes. Instead, we measure an upper limit on the stellar masses of our LRDs by fitting only the rest-UV portions of their spectra with a \textit{stellar-only} model, and we find substantial differences compared to models which are fitted to the full spectrum.
\end{itemize}

Finally, we use these results to examine the \mbhmstar\ relation of high-redshift non-LRD AGN over cosmic time. We find that the observed relation seems to evolve rapidly, with AGN at \z\,$\gtrsim$\,$4.5$ exhibiting elevated \mbhmstar\ ratios and those at \z\,$\lesssim$\,$3.5$ lying near local relations. Without a full accounting of all observational biases and systematic uncertainties, we cannot rule out selection effects as the cause of at least part of this evolution. However, this change in the observed relation occurs over a period of just $\sim$\,500\,Myr and is entirely driven by a shift in the inferred \mstar\ estimates of AGN rather than by an evolving \mbh\ distribution. If this reflects a rapid evolution of the \textit{intrinsic} \mbhmstar\ relation, this would imply that BHs at early times grew rapidly, outpacing the buildup of their hosts, with rapid galaxy assembly occurring at later times to return them to the local relation. This physical picture is consistent with the predictions of some theoretical BH growth models. However, further work will be required to determine the true physical origin of this observed trend.

\begin{acknowledgements}
Authors from UT Austin acknowledge that we work at an institution that sits on the Indigenous lands of Turtle Island, the ancestral name for what now is called North America. Moreover, we would like to acknowledge the Alabama-Coushatta, Caddo, Carrizo/Comecrudo, Coahuiltecan, Comanche, Kickapoo, Lipan Apache, Tonkawa and Ysleta Del Sur Pueblo, and all the American Indian and Indigenous Peoples and communities who have been or have become a part of these lands and territories in Texas.

A.R.G. acknowledges support from the National Science Foundation through the NSF Graduate Research Fellowship Program. SLF and AJT acknowledge support from the University of Texas at Austin, and NASA via STScI award JWST-GO-6368. AVG acknowledges support from the Spanish grant PID2022-138560NB-I00, funded by MCIN/AEI/10.13039/501100011033/FEDER, EU.

The \jwst\ data used in this work were accessed from the Mikulski Archive for Space Telescopes (MAST) at the Space Telescope Science Institute, and can be accessed at doi:\href{https://doi.org/10.17909/0eta-ym85}{10.17909/0eta-ym85}

\end{acknowledgements}

\facilities{HST (ACS), JWST (NIRCam, NIRSpec)}
\software{\texttt{BEAGLE}/\texttt{PYP-BEAGLE} \citep{chevallard_modelling_2016}, \texttt{SciPy} \citep{virtanen_scipy_2020}, \texttt{NumPy} \citep{harris_array_2020}, \texttt{Astropy} \citep{astropycollaboration_astropy_2022}, \texttt{Emcee} \citep{foreman-mackey_emcee_2019}, \texttt{IPython} \citep{perez_ipython_2007}, \texttt{Matplotlib} \citep{hunter_matplotlib_2007}}

\bibliography{2026-05-20}{}
\bibliographystyle{aasjournalv7}

\appendix

\section{BL AGN Sample Information Tables}\label{apdx:b}

\startlongtable
\begin{deluxetable}{lccccccccc}
\tablewidth{0pt}
\tablecaption{BL AGN sample information.}
\tablehead{
  Source ID & R.A.\ [deg] & Decl.\ [deg] & \z\ & LRD & $\mathrm{log}($$M_{\rm BH}/\mathrm{M_{\odot}})$ & \nodata & $\mathrm{log}$($L_{\rm acc}$) & $\mathrm{log}(f_{\rm AGN})$ & $\hat{\tau}_{\rm V\!,\,cont}$)
}
\startdata
CEERS-397 & 214.836197 & 52.882693 & 6.000 & 0 & $7.60^{+0.23}_{-0.18}$ & \nodata & $45.78^{+0.03}_{-0.04}$ & $1.72^{+0.04}_{-0.06}$ & $0.04^{+0.01}_{-0.01}$ \\
  CEERS-746 & 214.809142 & 52.868484 & 5.623 & 1 & $7.29^{+0.16}_{-0.14}$ & \nodata & \nodata & \nodata & \nodata \\
  RUBIES-EGS-920 & 215.052344 & 52.884268 & 3.616 & 0 & $6.33^{+0.36}_{-0.20}$ & \nodata & $44.15^{+0.08}_{-0.10}$ & $-0.59^{+0.40}_{-1.50}$ & $0.24^{+0.45}_{-0.13}$ \\
  RUBIES-UDS-5496 & 34.405872 & -5.312951 & 3.655 & 0 & $6.99^{+0.17}_{-0.17}$ & \nodata & $44.62^{+0.07}_{-0.07}$ & $1.55^{+0.06}_{-0.07}$ & $4.66^{+0.68}_{-0.69}$ \\
  RUBIES-EGS-6411 & 215.109185 & 52.939770 & 4.880 & 0 & $6.49^{+0.16}_{-0.16}$ & \nodata & $43.92^{+0.06}_{-0.07}$ & \nodata & \nodata \\
  RUBIES-UDS-8895 & 34.363041 & -5.306108 & 3.982 & 0 & $7.37^{+0.05}_{-0.05}$ & \nodata & $44.88^{+0.03}_{-0.03}$ & $2.21^{+0.08}_{-0.09}$ & $1.61^{+0.05}_{-0.08}$ \\
  RUBIES-UDS-10036 & 34.381671 & -5.303742 & 3.806 & 1 & $6.73^{+0.08}_{-0.07}$ & \nodata & \nodata & \nodata & \nodata \\
  RUBIES-UDS-11721 & 34.411039 & -5.300780 & 3.978 & 0 & $7.24^{+0.16}_{-0.16}$ & \nodata & $45.11^{+0.06}_{-0.08}$ & $1.69^{+0.11}_{-0.12}$ & $1.28^{+0.11}_{-0.09}$ \\
  RUBIES-EGS-13872 & 215.132933 & 52.970705 & 5.262 & 0 & $6.74^{+0.23}_{-0.28}$ & \nodata & $43.72^{+0.16}_{-0.18}$ & $0.95^{+0.05}_{-0.04}$ & $2.44^{+0.23}_{-0.18}$ \\
  RUBIES-UDS-16053 & 34.367104 & -5.293524 & 3.952 & 1 & $6.71^{+0.13}_{-0.13}$ & \nodata & \nodata & \nodata & \nodata \\
  RUBIES-EGS-17146 & 214.949482 & 52.845415 & 5.000 & 0 & $6.60^{+0.25}_{-0.15}$ & \nodata & $44.17^{+0.11}_{-0.16}$ & $0.67^{+0.16}_{-0.14}$ & $0.40^{+0.06}_{-0.06}$ \\
  RUBIES-UDS-18302 & 34.233628 & -5.283850 & 3.698 & 0 & $7.01^{+0.36}_{-0.41}$ & \nodata & $43.15^{+0.20}_{-0.11}$ & \nodata & \nodata \\
  RUBIES-EGS-19174 & 214.860840 & 52.784773 & 3.774 & 0 & $6.62^{+0.20}_{-0.16}$ & \nodata & $44.39^{+0.08}_{-0.07}$ & $1.21^{+0.03}_{-0.03}$ & $1.45^{+0.06}_{-0.05}$ \\
  RUBIES-UDS-19484 & 34.232426 & -5.280654 & 4.656 & 1 & $6.62^{+0.19}_{-0.22}$ & \nodata & \nodata & \nodata & \nodata \\
  RUBIES-UDS-19521 & 34.383672 & -5.287732 & 5.669 & 0 & $7.18^{+0.22}_{-0.26}$ & \nodata & $44.36^{+0.32}_{-0.04}$ & $0.85^{+0.13}_{-0.12}$ & $3.17^{+0.53}_{-0.45}$ \\
  RUBIES-UDS-21944 & 34.469218 & -5.283563 & 3.526 & 0 & $6.96^{+0.11}_{-0.13}$ & \nodata & $44.10^{+0.09}_{-0.11}$ & $1.23^{+0.04}_{-0.03}$ & $1.27^{+0.05}_{-0.05}$ \\
  RUBIES-EGS-28812 & 214.924149 & 52.849050 & 4.223 & 0 & $7.35^{+0.04}_{-0.04}$ & \nodata & $44.67^{+0.01}_{-0.01}$ & $1.06^{+0.02}_{-0.04}$ & $0.88^{+0.02}_{-0.01}$ \\
  RUBIES-EGS-29489 & 215.022071 & 52.920786 & 4.543 & 1 & $7.39^{+0.08}_{-0.09}$ & \nodata & \nodata & \nodata & \nodata \\
  RUBIES-UDS-29813 & 34.453355 & -5.270717 & 5.440 & 1 & $7.39^{+0.09}_{-0.09}$ & \nodata & \nodata & \nodata & \nodata \\
  RUBIES-UDS-31747 & 34.223757 & -5.260245 & 4.130 & 1 & $7.17^{+0.05}_{-0.05}$ & \nodata & \nodata & \nodata & \nodata \\
  RUBIES-EGS-37032 & 214.849388 & 52.811824 & 3.850 & 1 & $7.29^{+0.15}_{-0.13}$ & \nodata & \nodata & \nodata & \nodata \\
  RUBIES-EGS-37124 & 214.990977 & 52.916524 & 5.682 & 1 & $7.18^{+0.12}_{-0.11}$ & \nodata & \nodata & \nodata & \nodata \\
  RUBIES-EGS-42046 & 214.795368 & 52.788847 & 5.279 & 1 & $8.30^{+0.02}_{-0.02}$ & \nodata & \nodata & \nodata & \nodata \\
  RUBIES-EGS-42232 & 214.886792 & 52.855381 & 4.954 & 1 & $7.67^{+0.04}_{-0.03}$ & \nodata & \nodata & \nodata & \nodata \\
  RUBIES-UDS-46566 & 34.293543 & -5.234588 & 3.538 & 0 & $6.67^{+0.05}_{-0.05}$ & \nodata & $45.67^{+0.03}_{-0.03}$ & $1.65^{+0.07}_{-0.10}$ & $0.34^{+0.01}_{-0.01}$ \\
  RUBIES-UDS-47509 & 34.264602 & -5.232586 & 5.673 & 1 & $7.54^{+0.07}_{-0.07}$ & \nodata & \nodata & \nodata & \nodata \\
  RUBIES-EGS-49140 & 214.892248 & 52.877410 & 6.686 & 1 & $8.26^{+0.02}_{-0.01}$ & \nodata & \nodata & \nodata & \nodata \\
  RUBIES-EGS-50052 & 214.823454 & 52.830277 & 5.240 & 0 & $7.58^{+0.03}_{-0.03}$ & \nodata & $44.63^{+0.05}_{-0.05}$ & $0.98^{+0.04}_{-0.04}$ & $2.77^{+0.21}_{-0.19}$ \\
  RUBIES-EGS-50812 & 214.845487 & 52.848281 & 3.518 & 0 & $6.77^{+0.14}_{-0.13}$ & \nodata & $44.55^{+0.05}_{-0.06}$ & $0.48^{+0.04}_{-0.04}$ & $2.72^{+0.40}_{-0.29}$ \\
  \nodata & \nodata & \nodata & \nodata & \nodata & \nodata & \nodata & \nodata & \nodata & \nodata \\
\enddata
\tablecomments{Columns: 
(1) Source ID;
(2) Right ascension;
(3) Declination;
(4) Redshift;
(5) LRD flag;
(6) $\mathrm{log}(M_{\rm BH}/\mathrm{M_{\odot}})$;
(7) $\mathrm{log}(M_{*}/\mathrm{M_{\odot}})$;
(8) $\mathrm{log}(Z_{*}/\mathrm{Z}_{\odot})$;
(9) $\mathrm{log}(\tau_{\rm SFR}/\mathrm{yr})$;
(10) $\mathrm{log}(t_{\rm max}/\mathrm{yr})$;
(11) $\mathrm{log}(\mathrm{SFR}/\mathrm{M}_{\odot}\,\mathrm{yr}^{-1})$;
(12) $\mathrm{log}\,U^{\rm HII}$;
(13) $\hat{\tau}_{\rm V}$;
(14) $\mathrm{log}(Z_{\rm gas}^{\rm NLR}/\mathrm{Z}_{\odot})$;
(15) $\mathrm{log}\,U^{\rm NLR}$;
(16) $\mathrm{log}(L_{\rm acc})$;
(17) $\mathrm{log}(f_{\rm AGN})$;
(18) $\hat{\tau}_{\rm V\!,\,cont}$.
Best-fit parameters listed for each source are the median and 16/84th percentile values from the posterior distributions of the fiducial model. Since we define the \textit{stellar-only} fit to the rest-UV spectrum of each LRD as its fiducial model, AGN parameters are not reported for any LRDs. A machine-readable version of this table can be accessed at \href{https://zenodo.org/records/20419666}{https://zenodo.org/records/20419666}. }
\end{deluxetable}
\clearpage

\section{Additional Figures}

\begin{figure}[h!]
    \centering
    \includegraphics[width=0.96\linewidth]{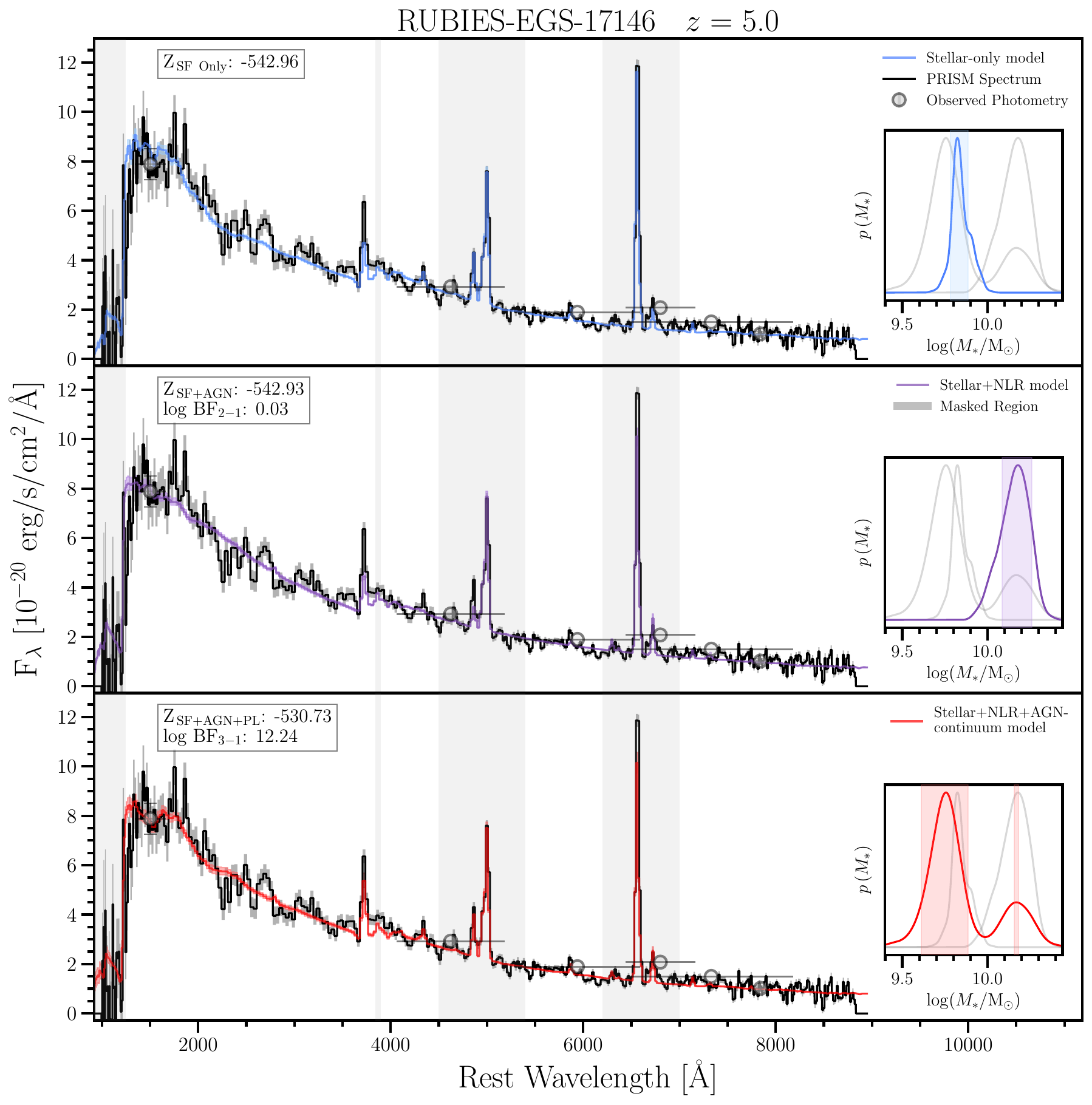}
    \caption{Comparison of the best-fit \textit{stellar-only} (top), \textit{stellar+NLR} (middle), and \textit{stellar+NLR+AGN-continuum} (bottom) model spectra to the observed PRISM spectrum of an example BL AGN in our sample. Inset plots show the \mstar\ posterior distributions for each model, and text in bottom two panels reports the BF for the \textit{stellar+NLR} and \textit{stellar+NLR+AGN-continuum} model, respectively, computed relative to the \textit{stellar-only} model. Note that we only fit the narrow component of \Ha, so the best-fit spectrum may not visually match the observed \Ha\ emission line even in a well-fitting model. The complete figure set for all sources in our BL AGN sample is available at \href{https://zenodo.org/records/20419666}{https://zenodo.org/records/20419666}.}
    \label{fig:17146-comparison}
\end{figure}

\begin{figure}[h!]
    \centering
    \includegraphics[width=\linewidth]{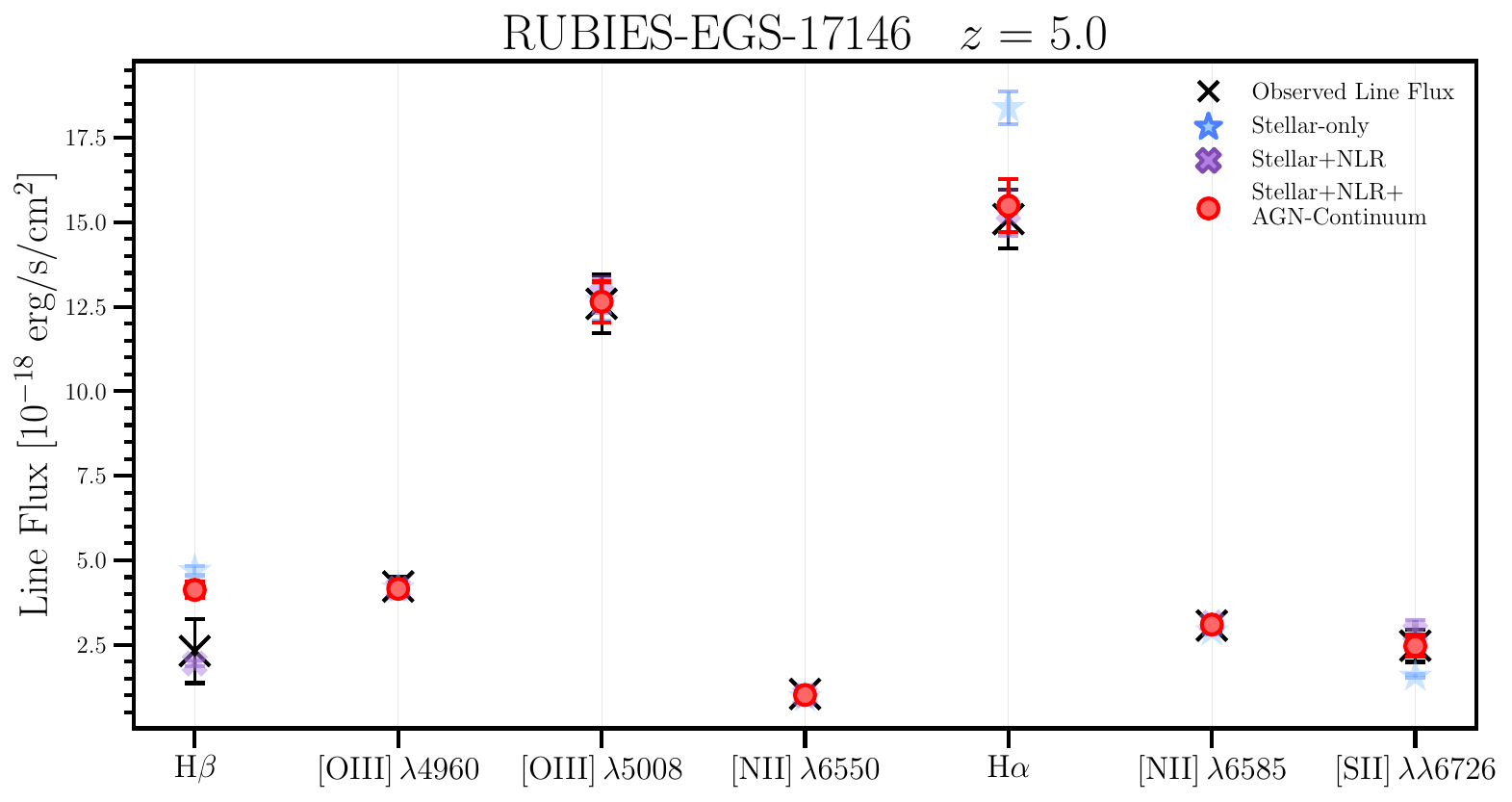}
    \caption{Comparison of the best-fit \textit{stellar-only}, \textit{stellar+NLR}, and \textit{stellar+NLR+AGN-continuum} model emission line fluxes to the line fluxes measured from the observed PRISM spectrum (applying ratios from a G395M decomposition) for an example BL AGN in our sample. The best-fit model fluxes (as measured by our BF comparison) are shown in full opacity, and results from the remaining fits are indicated by partially transparent markers. The complete figure set for all sources in our BL AGN sample is available at \href{https://zenodo.org/records/20419666}{https://zenodo.org/records/20419666}.}
    \label{fig:17146-spec-indices}
\end{figure}    

\begin{figure}[h!]
    \centering
    \includegraphics[width=\linewidth]{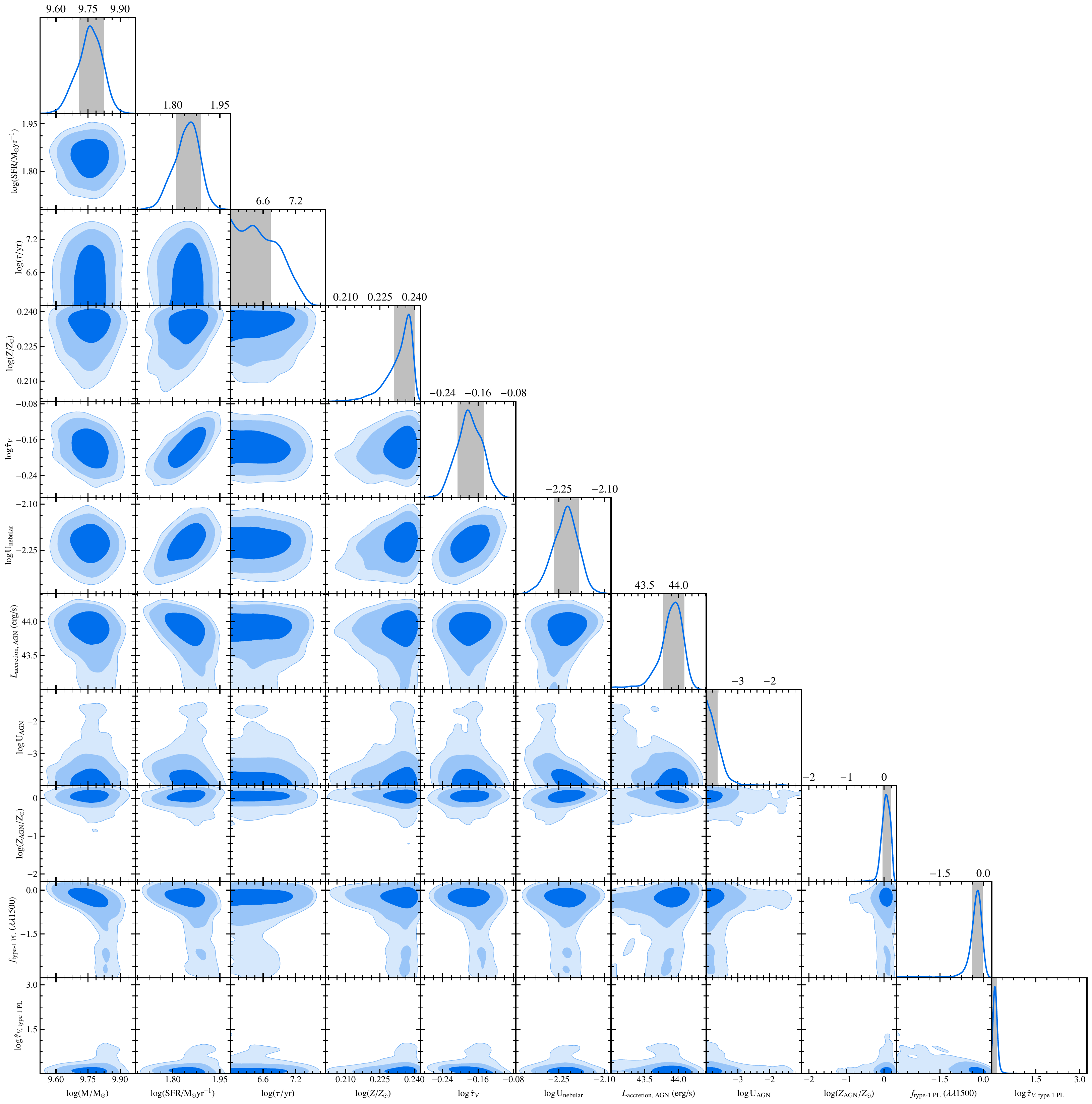}
    \caption{``Corner'' plot generated using the \textsc{PYP-BEAGLE} Python package showing posterior distributions for fitted parameters and their co-variances in the \textit{stellar+NLR+AGN-continuum} model fit to an example source (RUBIES-EGS-17146). The complete figure set for all sources in our BL AGN sample (and for each model type) is available at \href{https://zenodo.org/records/20419666}{https://zenodo.org/records/20419666}.}
    \label{fig:17146-triangle-plot}
\end{figure}
\clearpage

\section{Literature Sample}\label{apdx:c}
We compiled a comprehensive sample of BL AGN from all papers that we are aware of which report \mbh\ and \mstar\ estimates for spectroscopically-confirmed BL AGN with \jwst\ observations at 3\,$<$\,\z\,$<$\,7.5. Re-deriving all measured quantities is beyond the scope of this work, as sources in this sample are identified using a variety of \jwst\ instruments and configurations; as we mainly use this sample for qualitative analyses, we instead directly adopt the \mbh\ and \mstar\ estimates (and LRD classifications) reported in each work. To identify duplicate entries (sources which are analyzed in multiple papers), we perform a 1'' internal cross-match within the full literature sample. Among each group of duplicates, we select the measurement with the lowest error reported in \mstar\ for our final literature sample. To remove any objects which are independently analyzed in this work, we finally perform a 1'' cross-match to our BL AGN sample and remove any matched sources.

\startlongtable
\begin{deluxetable}{llcccccc}
\tablewidth{0pt}
\tablecaption{Source information for the literature BL AGN sample.}
\tablehead{
  Source ID & R.A. & Decl. & \z\ & $\mathrm{log}($$M_{\rm BH}/\mathrm{M_{\odot}}$) & $\mathrm{log}($$M_{\rm*}/\mathrm{M_{\odot}}$) & LRD & Reference
}
\startdata
  COS-6696 & 150.125824 & 2.389711 & 7.037 & $7.28^{+0.27}_{-0.20}$ & $8.49^{+0.18}_{-0.17}$ & 1 & \citetalias{akins_strong_2025} \\
  RUBIES-37427 & 34.504940 & -5.257980 & 6.958 & $7.42^{+0.04}_{-0.05}$ & $8.50^{+0.20}_{-0.10}$ & 0 & \citetalias{arevalo-gonzalez_agn_2026} \\
  JADES-167639 & 53.091880 & -27.880700 & 5.555 & $7.70^{+0.04}_{-0.05}$ & $8.28^{+0.04}_{-0.04}$ & 0 & \citetalias{arevalo-gonzalez_agn_2026} \\
  CAPERS-52661 & 150.141800 & 2.193836 & 6.764 & $7.81^{+0.04}_{-0.05}$ & $9.20^{+0.10}_{-0.10}$ & 0 & \citetalias{arevalo-gonzalez_agn_2026} \\
  JADES-38147 & 189.270700 & 62.148420 & 5.878 & $7.84^{+0.04}_{-0.05}$ & $7.90^{+0.50}_{-0.10}$ & 0 & \citetalias{arevalo-gonzalez_agn_2026} \\
  \nodata & \nodata & \nodata & \nodata & \nodata & \nodata & \nodata & \nodata
\enddata
\tablecomments{(1) \citealt{akins_strong_2025}; (2) \citealt{arevalo-gonzalez_agn_2026}; Additional sources from the following publications are included: \citet{arevalo-gonzalez_agn_2026, carnall_massive_2023, degraaff_remarkable_2025, deugenio_blackthunder_2025, fei_glimpse_2025, harikane_jwst_2023, jones_m_rm_2025, juodzbalis_jades_2026, kim_jwst_2025, kiyota_comprehensive_2025, kokorev_deepest_2025, li_cosmos3d_2025, li_dichotomy_2026, li_prevalent_2025, maiolino_jades_2024, marshall_ganifs_2025, marshall_jwsts_2025, parlanti_ganifs_2024, schindler_little_2025, stone_detection_2023, stone_undermassive_2024, ubler_blackthunder_2025, ubler_ganifs_2023, ubler_ganifs_2024a, yue_eiger_2024, zhang_abundant_2026}. A machine-readable version of this table can be accessed at \href{https://zenodo.org/records/20419666}{https://zenodo.org/records/20419666}.}
\end{deluxetable}

\end{document}